\documentclass[apjl]{emulateapj}

\newcommand {\INTEGRAL}{{\it INTEGRAL }}

\newcommand {\early}{{\bf early }}
\newcommand {\late}{{\bf late }}
\newcommand{\W}[1]{\texttt{W#1}}
\newcommand{\HED}[1]{\texttt{HED#1}}
\newcommand{\BALL}[1]{\texttt{#1Dbbal}}
\newcommand{\DD}[1]{\texttt{DD#1}}
\newcommand{\PAR}[1]{\texttt{#1PAR}}
\newcommand{\DDTe}{\texttt{DDTe }}
\newcommand {\tem}{{\texttt{TEM} }}
\newcommand {\DETO}{{\texttt{DETO} }}
\newcommand {\bc}{\begin {center}}
\newcommand {\ec}{\end {center}}
\newcommand {\be}{\begin {equation}}
\newcommand {\ee}{\end {equation}}
\newcommand {\eqref}[1]{equation (\ref{#1})}

\def\deg{$^{\circ}~$}

\shorttitle{}
\shortauthors{Churazov et al.}

\begin{document}
\title{Gamma-rays from Type Ia supernova SN2014J}
%\author{}
\author{E.~Churazov\altaffilmark{1,2}, R.~Sunyaev\altaffilmark{1,2},
  J.~Isern$^{3}$, I.~Bikmaev$^{4,5}$, E.~Bravo$^{6}$, N.~Chugai$^{7}$,
  S.~Grebenev$^{1}$, P.~Jean$^{8,9}$, J.~Kn\"odlseder$^{8,9}$,
  F.~Lebrun$^{10}$, E.~Kuulkers$^{11}$}

\affil{$^1$Space Research Institute (IKI), Profsouznaya 84/32, Moscow
  117997, Russia
}
\affil{$^2$Max
Planck Institute for Astrophysics, Karl-Schwarzschild-Strasse 1, 85741 Garching, Germany
}
\affil{$^3$
Institut for Space Sciences (ICE-CSIC/IEEC), 08193 Bellaterra, Spain
}
\affil{$^4$
Kazan Federal University (KFU), Kremlevskaya Str., 18, Kazan, Russia
}
\affil{$^5$
Academy of Sciences of Tatarstan, Bauman Str., 20, Kazan, Russia
}
\affil{$^6$E.T.S.A.V., Univ. Politecnica de Catalunya, Carrer Pere Serra 1-15, 08173 Sant Cugat
del Valles, Spain
}
\affil{$^7$Institute of Astronomy of the Russian Academy of Sciences,
  48 Pyatnitskaya St. 119017, Moscow, Russia}

\affil{
$^8$Universit\'e de Toulouse; UPS-OMP; IRAP;  Toulouse, France}

\affil{$^9$CNRS; IRAP; 9 Av. colonel Roche, BP 44346, F-31028 Toulouse cedex
  4, France
}

%\affil{$^{9}$LUPM, Universit\'e Montpellier 2, CNRS/IN2P3, CC 72,
%  Place Eug\`ene Bataillon, F-34095 Montpellier Cedex 5, France
%}

\affil{$^{10}$APC, Univ Paris Diderot, CNRS/IN2P3, CEA/Irfu, Obs de
  Paris, Sorbonne Paris Cit\'e, France}

\affil{$^{11}$ European Space Astronomy Centre (ESA/ESAC), Science Operations Department, P.O. Box 78, 28691 Villanueva de la Ca\~nada, Madrid, Spain}

\begin{abstract}
The whole set of \INTEGRAL observations of type Ia supernova SN2014J,
covering the period 19-162 days after the explosion has being analyzed. For
spectral fitting the data are split into \early and \late periods
covering days 19-35 and 50-162, respectively, optimized for $^{56}$Ni
and $^{56}$Co lines. As expected for the \early period much of the
gamma-ray signal is confined to energies below $\sim$200 keV, while
for the \late period it is most strong above 400 keV. In particular, in the \late period $^{56}$Co lines at 847 and 1248 keV are detected at 4.7 and 4.3 $\sigma$ respectively.
The lightcurves in several representative energy
bands are calculated for the entire period. The resulting spectra and
lightcurves are compared with a subset of models. We confirm our
previous finding that the gamma-ray data are broadly consistent with
the expectations for canonical 1D models, such as delayed detonation or
deflagration models for a near-Chandrasekhar mass WD. Late optical
spectra (day 136 after the explosion) show rather symmetric Co and Fe lines
profiles, suggesting that unless the viewing angle is special, the
distribution of radioactive elements is symmetric in the ejecta.  
\end{abstract}
\keywords{}

\section{Introduction}
\label{sec:into}

A Type Ia supernova is believed to be a thermonuclear explosion of a
carbon-oxygen (CO) white dwarf \citep{1960ApJ...132..565H} in a
binary system, \citep[see, e.g.,][for a
  review]{2000ARA&A..38..191H,2005AstL...31..528I}. Most popular scenarios of the
explosion include (i) a gradual increase of the mass towards
Chandrasekhar limit \citep[e.g.,][]{1973ApJ...186.1007W}, (ii) a
merger/collision of two WDs
\citep[e.g.,][]{1984ApJS...54..335I,1984ApJ...277..355W,2013ApJ...778L..37K},
(iii) an initial explosion at the surface of the sub-Chandrasekhar WD,
which triggers subsequent explosion of the bulk of the material
\citep[e.g.,][]{1977PASJ...29..765N,1996ApJ...457..500H}. In all
scenarios a thermonuclear runaway converts substantial fraction of CO
mass into iron-group elements and the released energy powers the
explosion itself. The optical light of the supernova is in turn
powered by the decay of radioactive elements, synthesized during
explosion. For the first year since the explosion the decay chain of
$^{56}$Ni$\rightarrow^{56}$Co$\rightarrow^{56}$Fe is of prime
importance. As long as the expanding ejecta are optically thick for
gamma-rays the bulk of the decay energy is thermalized and is
re-emitted in the UV, optical and IR band. After several tens of days
the ejecta become optically thin for gamma-rays making SNIa a powerful
source of gamma photons.

Here we report the results of \INTEGRAL observations of
SN2014J covering a period from $\sim$16 to $\sim$162 days since the
explosion.  

The analysis of the SN2014J data obtained by \INTEGRAL has been
reported in \citet{2014Natur.512..406C} (days $\sim$50-100 since
explosion), \citet{2014Sci...345.1162D} (days $\sim$16-19),
\citet{isern} (days $\sim$16-35), see also
\citet{2015A&A...574A..72D}. Despite of the proximity, SN2014J in
gamma-rays is an extremely faint source and the expected signal is
below 1\% of the background. This makes the results sensitive to the
adopted procedure of the background handling by different groups and
lead to tension between some results. Here we have combined all \INTEGRAL
data and uniformly process them using the same procedure as in
\citet{2014Natur.512..406C}.  The resulting spectra and light-curves
are compared with the predictions of basic type Ia models.

Current state-of-the-art 3D simulations of type Ia explosions
\citep[e.g.,][]{2013MNRAS.429.1156S,2014MNRAS.438.1762F,2014ApJ...785..105M}
lead to a complicated distribution of burning products in the ejecta
and introduce a viewing angle dependence in the predicted gamma-ray
flux. However, the overall significance of the SN2014J detection in
gamma-rays by \INTEGRAL (see \S\ref{sec:observations} and
\S\ref{sec:results}) corresponds to $\sim 10$ s.t.d. This precludes a
very detail model-independent analysis.  We therefore took a
conservative approach of comparing the data with a subset of popular
1D SNIa models (see \S\ref{sec:models}), some of which were used in
\citet{2004ApJ...613.1101M} for assesment of SNIa gamma-ray
codes. While these models do not describe the full complexity of SNIa
ejecta, they can serve as useful indicators of the most basic
characteristics of the explosion, including the total mass of
radioactive nickel, total mass of the ejecta and the expansion
velocity.  We also verify (\S\ref{sec:tem_late}) if adding an extra
component, corresponding to a transparent clump of radioactive Ni, on
top of the best-fitting 1D model, significantly improves the fit.  In
\S\ref{sec:opt} we make several basic consistency checks of gamma-ray
and optical data, using optical observations taken
quasi-simultaneously with \INTEGRAL observations. Section
\ref{sec:conclusions} provides the summary of our results.

\section{SN2014J in M82}
\label{sec:sn2014j}
SN2014J in M82 was discovered \citep{2014CBET.3792....1F} on Jan. 21,
2014. The reconstructed
\citep{2014ApJ...783L..24Z,2015ApJ...799..106G} date of the explosion
is Jan. 14.75 UT with the uncertainty of order $\pm0.3$ days. At the
distance of M82 ($\sim3.5$ Mpc), this is the nearest SN Ia in several
decades. The proximity of the SN2014J triggered many follow-up
observations, including those by \INTEGRAL
\citep{2014ATel.5835....1K}.

The SN is located $\sim$1~kpc from the M82 nucleus and has a strong
($A_{V}\sim 2$) and complicated absorption in the UV-optical band
\citep[e.g.,][]{2014ApJ...784L..12G,2015ApJ...798...39M,2014ApJ...788L..21A,2014MNRAS.443.2887F,2014ApJ...792..106W,2015A&A...577A..53P,2015ApJ...805...74B,2014ApJ...795L...4K}.

From the light curves and spectra SN2014J appears to be a ``normal''
SNIa with no large mixing
\citep[e.g.,][]{2015ApJ...798...39M,2014MNRAS.445.4427A}, consistent
with the delayed-detonation models. Detection of stable Ni
\citep{2014ApJ...792..120F,2015ApJ...798...93T} in IR suggests high
density of the burning material \citep[see, e.g.,][]{1992ApJ...386L..13S}, characteristic for near-Chandrasekhar
WD.

Search in X-ray, radio and optical bands (including pre-supernova
observations of M82) didn't reveal any evidence for accretion onto the
WD before the explosion, any candidate for a companion star, or
compelling evidence for a large amount of circumbinary material,
implicitly supporting the DD scenario
`\citep{2014ApJ...790....3K,2014MNRAS.442.3400N,2014ApJ...790...52M,2014ApJ...792...38P},
although some SD scenarios are not excluded.

In gamma-rays the first detection of SN2014J in $^{56}$Co lines was reported
 about 50 days since the explosion \citep{2014ATel.5992....1C}. The
gamma-ray signal from SN2014J was also reported in the earlier phase
$\sim$16-35 days after the explosion \citep{2014ATel.6099....1I,2014Sci...345.1162D}.

Throughout the paper we adopt the distance to M82 (and to SN2014J) of
3.5 Mpc. The recent analysis by \citet{2014MNRAS.443.2887F} suggests the
distance of $3.27\pm0.2$ Mpc. This estimate is formally consistent with
the $D\sim 3.53\pm0.26$ Mpc from \citealt{2006Ap.....49....3K} and our
adopted value. Nevertheless, one should bear in mind that all fluxes
and normalizations of best-fitting models can be overestimated
(underestimated) by as much as $\sim20$\%.

The only other supernova sufficiently bright to allow for detailed study in gamma-rays from  $^{56}$Ni
and $^{56}$Co decay is the Type II SN1987A in Large Magellanic Cloud. In SN1987A the down-scattered hard X-ray continuum was first seen half a year after the explosion \citep{1987Natur.330..227S,1987Natur.330..230D,1990SvAL...16..171S}, while $\gamma$-ray lines of $^{56}$Co were detected several months later \citep{1988Natur.331..416M,1989Natur.339..122T}. While SN2014J is more than 60 times further away from us than SN1987A, the larger amount of radioactive $^{56}$Ni and less massive/opaque ejecta in type Ia supernovae made the detection of gamma-rays from SN2014J possible. 

\section{\INTEGRAL Observations and basic data analysis}
\label{sec:observations}
\INTEGRAL is an ESA scientific mission dedicated to fine
spectroscopy and imaging of celestial $\gamma$-ray sources in the
energy range 15\,keV to 10\,MeV \citep{2003A&A...411L...1W}.

The \INTEGRAL data used here were accumulated during revolutions
1380-1386, 1391-1407 and
1419-1428\footnote{http://www.cosmos.esa.int/web/integral/schedule-information
},
corresponding to the period $\sim$16-162 days after the explosion.

In the analysis we follow the procedures described in
\citet{2014Natur.512..406C,2013A&A...552A..97I} and use the data of two instruments
SPI and ISGRI/IBIS on board \INTEGRAL.

\subsection{SPI}
\label{sec:spi}
SPI is a coded mask germanium spectrometer on board \INTEGRAL.
The instrument consists of 19 individual Ge detectors, has a field of
view of $\sim$30\deg (at zero response), an effective area $\sim
70$~cm$^2$ at 0.5 MeV and energy resolution of $\sim$2 keV
\citep{2003A&A...411L..63V,2003A&A...411L..91R}. Effective angular resolution of SPI is
$\sim$2\deg.  During SN2014J observations 15 out of 19 detectors were
operating, resulting in slightly reduced sensitivity and imaging
capabilities compared to initial configuration. 

Periods of very high and variable background due to solar flares and passage through radiation belts were
omitted from the analysis. In particular, based on the SPI
anti-coincidence system count-rates, the revolutions 1389 and 1390
were completely excluded, as well as parts of revolutions 1405, 1406,
1419, 1423 and 1426.  The data analysis follows the scheme
implemented for the analysis of the Galactic Center positron
annihilation emission \citep{2005MNRAS.357.1377C,2011MNRAS.411.1727C}. We used only ``single'' events \citep{2003A&A...411L..63V} and for
each detector, a linear relation between the energy and the channel
number was assumed and calibrated (separately for each orbit), using
the observed energies of background lines at ~198, 438, 584, 882,
1764, 1779, 2223 and 2754 keV.

The flux of the supernova $S(E)$ at energy $E$ and the background
rates in individual detectors $B_i(E,t)$ were derived from a simple model
of the observed rates $D_i(E,t)$ in individual SPI detectors, where $i$ is
the detector number and $t$ is the time of observation with a typical
exposure of 2000 s: 
\begin{eqnarray}
D_i(E,t)\approx S(E)\times R_i(E,t)+B_i(E,t).  
\end{eqnarray}
Here $R_i(E,t)$ is the effective area for the $i$-th detector, as seen from the
source position in a given observation. The background rate is assumed to be linearly proportional to
the Ge detectors' saturated event rate $G_{Sat}(t)$ above 8 MeV, averaged
over all detectors, i.e. $B_i(E,t)=\beta_i(E)G_{Sat}(t)+C_i(E)$, where
$C_i(E)$ does not depend on time.
The coefficients $S(E)$,$\beta_i(E)$ and $C_i(E)$ are free parameters of the model and are obtained by
minimizing $\chi^2$ for the entire data set. Even though the number of
counts in individual exposures is low, it is still possible to use a
plain $\chi^2$ approach as long as the errors are estimated using the mean
count rate and the total number of counts in the entire data set is
large \citep{1996ApJ...471..673C}. The linear nature of the model
allows for straightforward estimation of statistical errors.

\begin{figure*}
\begin{center}
\includegraphics[trim = 0 50mm 0 90mm,scale=0.7,clip]{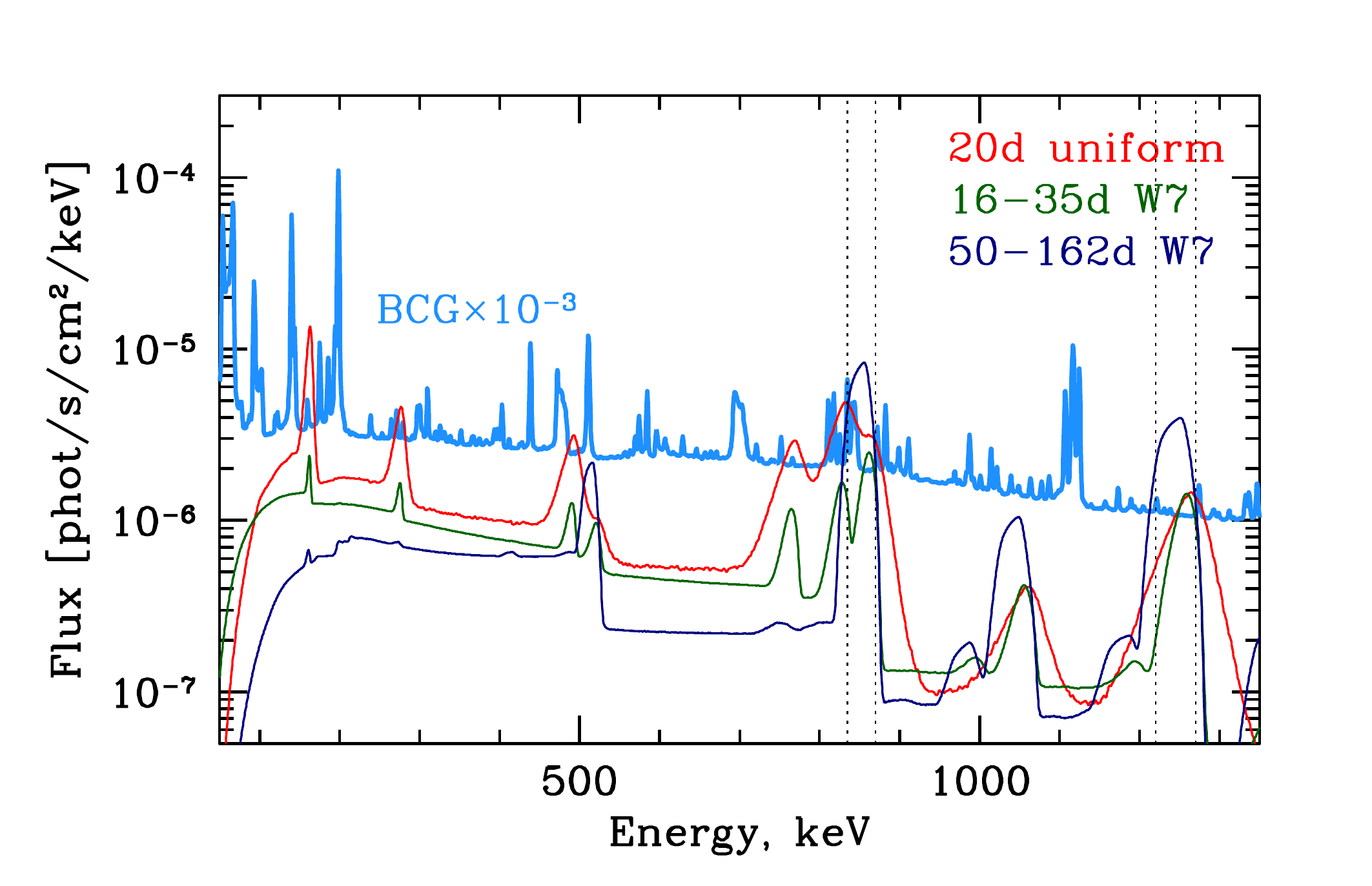}
\end{center}
\caption{SPI quiescent background in comparison with the
  representative model spectra. SPI background is multiplied by a
  factor $10^{-3}$. Green and blue lines correspond to the \W7 model
  \citep{1984ApJ...286..644N} averaged over the \early and \late periods (see \S\ref{sec:periods}),
  respectively. Red line shows the model the \PAR3 model from \citet{2014Natur.512..406C} for
  day 20 since the explosion. In this model all elements, including
  radioactive isotopes, are mixed uniformly over entire
  ejecta. The robust prediction of all plausible models is  the
  presence of two $^{56}$Co line at 847 and 1238 keV during the late phase. Vertical lines
  show two energy bands used for making images. The ``cleanest'' SPI
  background is near the 1238 keV line, where no strong instrumental
  lines are present.
 \label{fig:background}}
\end{figure*}

Despite its proximity, SN2014J is still an extremely faint source in
$\gamma$-rays. Fig.\ref{fig:background} shows the comparison of the
quiescent SPI background, scaled down by a factor of $10^3$ with a
sample of representative models. Two models labeled ``20d uniform''
and ``16-35d W7'' show the models for the early period of SN2014J
observations. The former model is based on 
the best-fitting \PAR3 model to the SN
spectra recorded between 50-100 days after explosion
\citep{2014Natur.512..406C}, recalculated for day 20. The model assumes
uniform mixing of all elements, including the radioactive $^{56}$Ni,
across the ejecta. This model at day 20 produces prominent $^{56}$Ni
lines near 158 keV and 812 keV. The latter model (\W7, see
\S\ref{sec:models}) averaged over period 16-35 days does not include
mixing and it produces much fainter lines. Finally the ``50-162d W7''
model corresponds to later observations. The most prominent features
of this model are the $^{56}$Co lines at 847 and 1238 keV. Among all
these features the $^{56}$Co line at 1238 keV is located in the least
complicated portion of the background spectrum.
 
The spectral redistribution matrix accounts for the instrumental line
broadening estimated from the data, accumulated during SN2014J
observations. We parametrize the energy resolution as a Gaussian with
the energy dependent width 
\begin{eqnarray}
\sigma_i\approx0.94~(E_{line}/500)^{0.115}~{\rm keV}.
\label{eq:sigma_i}
\end{eqnarray}

\begin{figure}
\begin{center}
\includegraphics[trim= 0cm 5cm 0cm 2cm,
  width=1\textwidth,clip=t,angle=0.,scale=0.49]{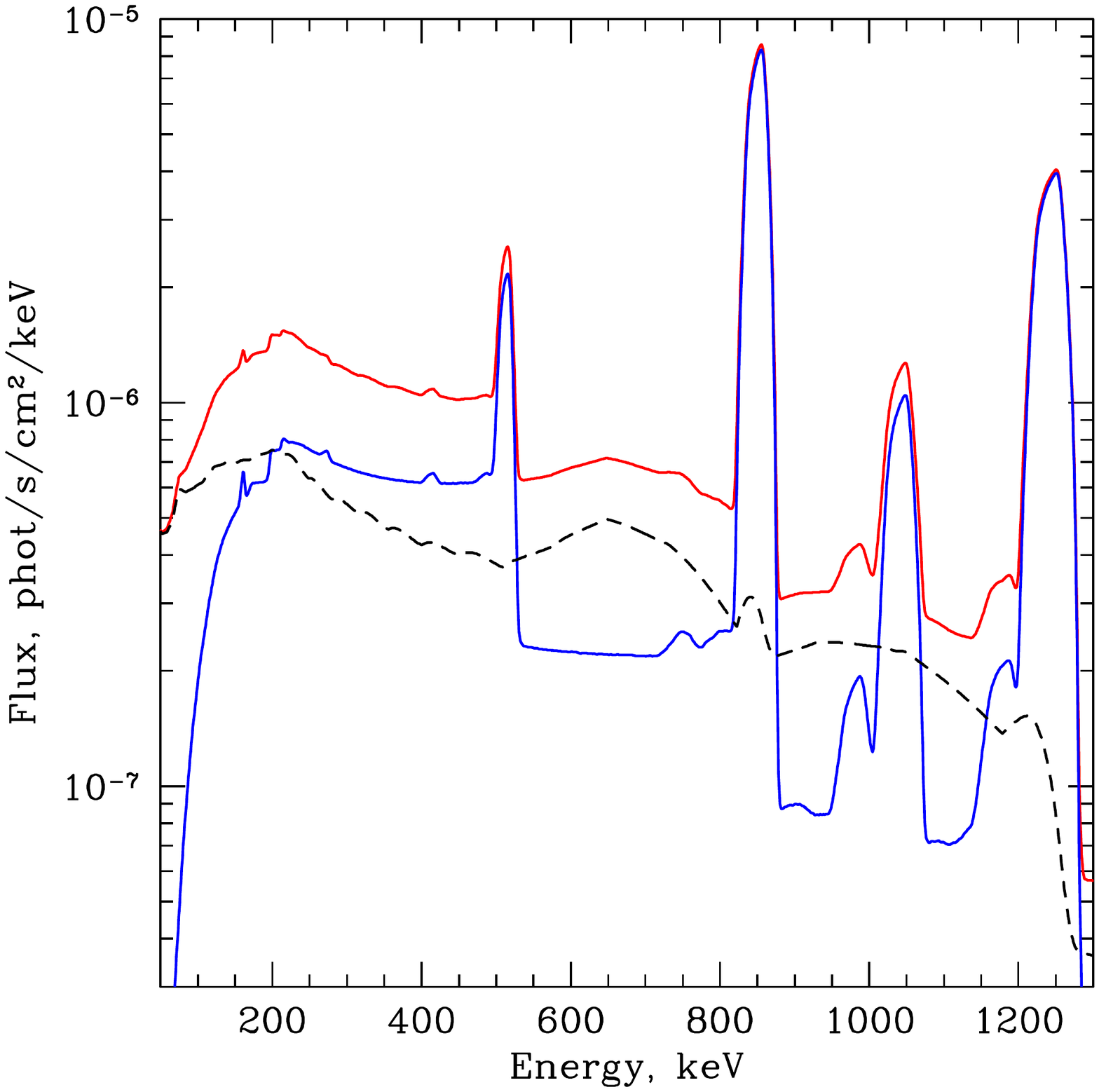}
\end{center}
\caption{Estimated contribution of the off-diagonal terms in the SPI
  spectral response to the SN spectrum. The blue line shows the
  predicted spectrum of the \W7 model for the \late period, convolved with a
  simplified (nearly diagonal) SPI response. In this approximation the
  instrumental broadening is parametrized as an energy dependent
  Gaussian with the width according to eq.\ref{eq:sigma_i}. The red line shows
  the same spectrum convolved with the response which includes
  estimated off-diagonal terms, caused by Compton scattering of
  incident photons in the detector and surrounding structures. The
  off-diagonal component alone is shown with the dashed black line.
  The off-diagonal terms create a long low-energy tails associated
  with gamma-ray lines. The impact on the brightest lines is small,
  while the continuum is strongly affected, especially at low
  energies. The model \W7 is averaged over the period 50-162 days after
  the explosion.
\label{fig:offdiag}}
\end{figure}

Compared to our previous analysis we amended the spectral
redistribution matrix of SPI by including low energy tails, associated
with the interactions (Compton scattering) of incoming photons inside
the detector and in the surrounding material. These photons are still
registered as single events in the SPI data, but their energies are
lower than the true incident energy. We used the results of
Monte-Carlo simulation of SPI energy/imaging response
\citep{2003A&A...411L..81S} and folded-in our procedure of spectrum
reconstruction described above. For steep spectra the account for low
energy tail results in a modest $\sim$10\% change in the spectrum
normalization, while for the very hard SN2014J spectrum it produces a
low energy tail which provides large contribution to the continuum,
while fluxes of narrow lines remain unaffected
(Fig.\ref{fig:offdiag}). With this response matrix the Crab Nebula
spectrum, observed by \INTEGRAL made between Feb 21 and 23, 2014, is
well described by a broken power law obtained by
\citet{2009ApJ...704...17J} for earlier Crab Nebula observations with
\INTEGRAL.

In our analysis we usually ignore the part of the spectrum at energies
higher than 1350 keV, since in the energy range between 1400 and 1700
keV the instrument suffers from the enhanced detector electronic
noise, while at even higher energies only weaker lines from $^{56}$Co
decay are expected (see Table~\ref{tab:tem} in \S\ref{sec:tem}).
  The convolution of the fiducial SNIa model (see \S\ref{sec:models})
  with the simulated SPI response \citep{2003A&A...411L..81S}
  confirmed that the contribution of high energy lines is negligible
  below 1350 keV, at least for ``single'' events considered here.

 The inspection of Fig.\ref{fig:background} shows that there is no
  chance to detect continuum in the SPI data for any of our fiducial
  models. E.g., for a 100 keV wide energy bin between 600 and 700 keV
  the expected $S/N$ after 4 Msec observation between days 50 and 162
  is $\sim 0.5 \sigma$. In the real data no evidence for significant
  continuum above 500 keV was found in the time-averaged spectra (see
  \S\ref{sec:spectra} below). As Fig.~\ref{fig:offdiag} the off-diagonal tail of the 847 and 1238 keV lines dominates over
  intrinsic SN continuum (see Fig.~\ref{fig:offdiag}), while the line
  shapes and fluxes are not affected. 

  In general, we consider the inclusion of the off-diagonal term in
  the response as an improvement compared to a pure diagonal
  response. We used this improved response throughout the paper and at the same time in \S\ref{sec:spectra} we consider several data sets,
  which include or exclude the SPI data below $\sim$400 keV. Inclusion
  of the low energy ($\la 400$ keV) data boosts the S/N, while the
  exclusion of these data (dominated by off-diagonal continuum) makes
  spectral fits less prone to possible uncertainties in the
  off-diagonal term calibration.
  
To verify the whole SPI pipeline, we have done an independent analysis
of the same data using the tools and procedures originally developed
and tuned for SN2011fe \citep[see][]{2013A&A...552A..97I}. This
analysis includes energy calibration, background modeling and the
background and source fluxes fitting. Verification of these steps is
important since the source (SN2014J) is very faint and even subtle
changes in the calibration might result in significant changes in the
source spectrum. The fluxes in the 835-870 keV band were derived using
these two independent pipelines for every revolution during SN2014J
observations. Comparing fluxes point by point, we have found very good
agreement, with the scatter well within statistical errors. The signal
from SN2014J is seen in both pipelines. No systematic trends of
deviations with the variations of the flux level are found.  We have
concluded that the results are fully consistent, within the
assumptions made on the background parameterization.

\subsection{ISGRI/IBIS}
\label{sec:isgri}

The primary imaging instrument inboard \INTEGRAL is IBIS \citep{2003A&A...411L.131U} - a
coded-mask aperture telescope with the CdTe-based detector ISGRI \citep{2003A&A...411L.141L}. It
has higher sensitivity to continuum emission than SPI in the 20-300
keV range\footnote{\texttt{http://www.cosmos.esa.int/web/integral/ao13}} and has a spatial resolution $\sim12'$. We note here, that
neither ISGRI, nor SPI can distinguish the emission of SN2014J from
the emission of any other source in M82. In particular, M82 hosts
two ultra-luminous and variable sources
\citep[e.g.][]{2014AstL...40...65S,2014Natur.514..202B} which 
contribute to the flux below $\sim 50$ keV. ISGRI however can easily
differentiate between M82 and M81, which are separated by $\sim30'$. The
energy resolution of ISGRI is $\sim$10\% at 100 keV. The ISGRI energy
calibration uses the procedure implemented in OSA 10.039. The images
in broad energy bands were reconstructed using a standard
mask/detector cross-correlation procedure, tuned to produce zero
signal on the sky if the count rate across the detector matches the
pattern expected from pure background, which was derived from the same
dataset by stacking detector images. The noise in the resulting images
is fully consistent with the expected level, determined by photon
counting statistics. The fluxes in broad bands were calibrated using
the 
Crab Nebula observations with \INTEGRAL made between Feb 21 and 23. The
\citet{2009ApJ...704...17J} model was assumed as a reference.

\subsection{Lightcurves, Spectra and Images}
\label{sec:periods}
\begin{figure*}
\begin{center}
\includegraphics[trim= 0cm 3cm 0cm 10cm,
  width=1\textwidth,clip=t,angle=0.,scale=0.99]{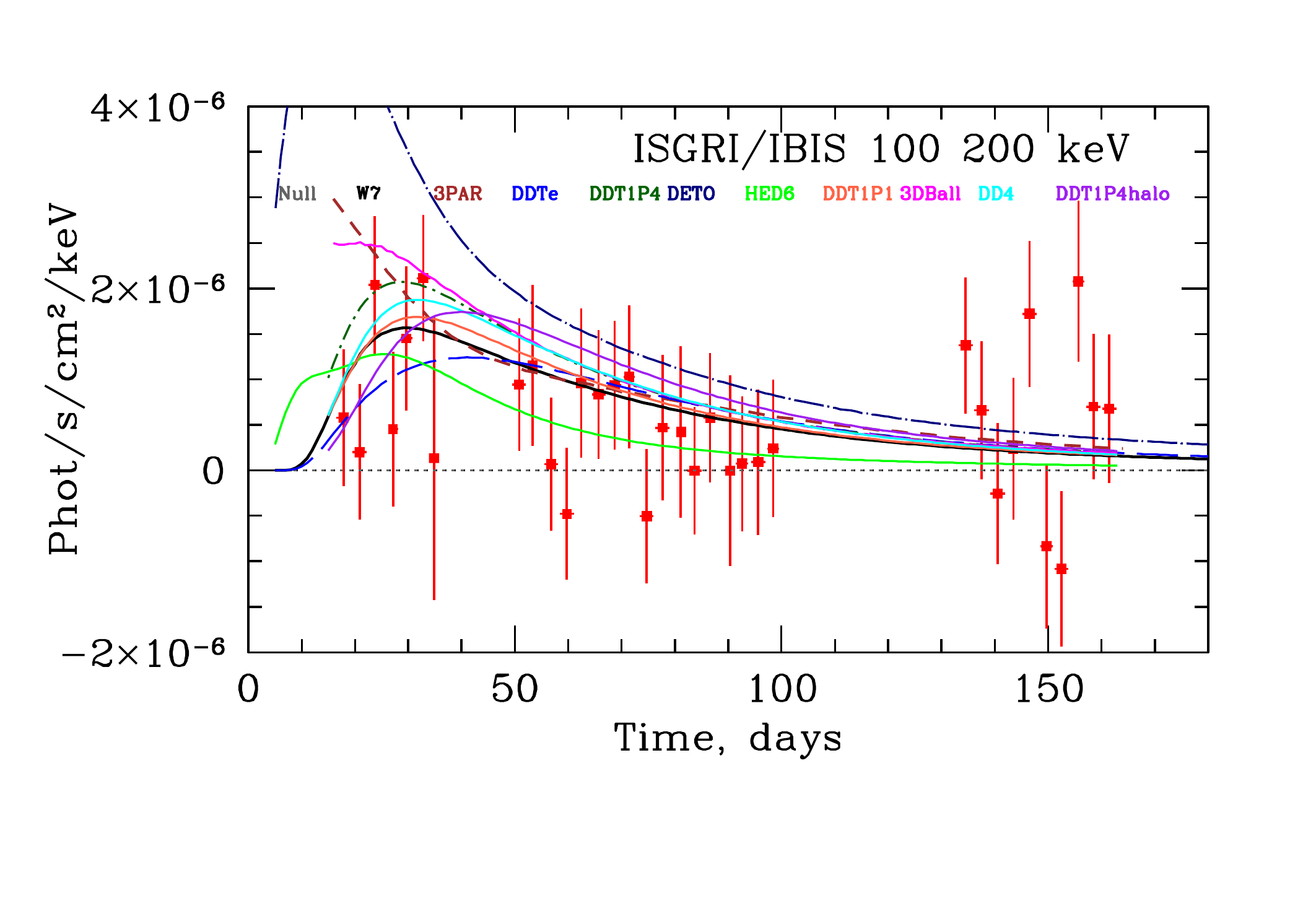}
\end{center}
\caption{ISGRI light curve in the 100-200 keV band. The S/N ratio in
    this band is expected to be the highest for the plausible
  models. The curves show the expected flux evolution for a
  set of models (see \S\ref{sec:models}). Color coding is explained in
  the legend.
\label{fig:lc_isgri}}
\end{figure*}

\begin{figure*}
\begin{center}
\includegraphics[trim= 0cm 5cm 0cm 2cm,
  width=1\textwidth,clip=t,angle=0.,scale=0.99]{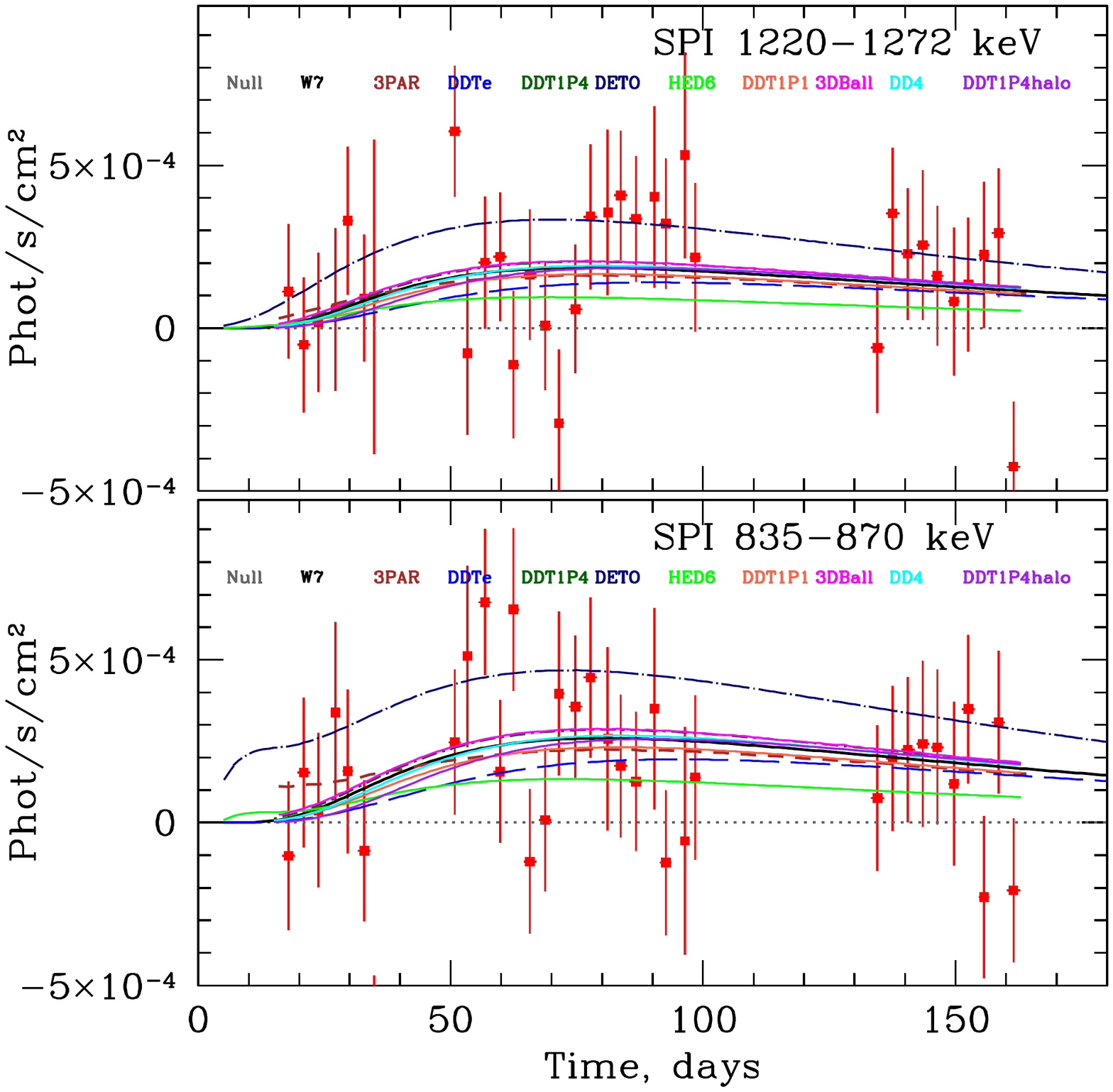}
\end{center}
\caption{The same as in Fig.\ref{fig:lc_isgri} for SPI data in two
  narrow bands near the brightest $^{56}$Co lines.
\label{fig:lc_spi}}
\end{figure*}

The lightcurves in several energy bands were generated using IGSRI and
SPI data. The time bins ($\sim$3 days each) correspond to individual
revolutions of the satellite. Finer time bins are not practical given
that the source is very faint. The lightcurves are shown in
Figs.\ref{fig:lc_isgri}-\ref{fig:lc_spi} together with a set of
representative models (see \S\ref{sec:models}). For the broad
100-200 keV band the conversion of the ISGRI flux using Crab spectrum
as a reference is not very accurate because of the difference in the
shape of the incident spectra. The conversion factor has been
recalculated using several representative SN models, resulting in a
modest $\sim$13\% correction factor, applied to the fluxes shown in Fig. \ref{fig:lc_isgri}.

In principle, the spectra can be extracted for any interval covered by
the observations, e.g., for individual revolutions, as is done above
for the lightcurves in several broad bands. For comparison of the
observed and predicted spectra we decided to split the data into two
intervals covering 16-35 and 50-162 days after the explosion,
respectively (see Table~\ref{tab:sets}). The gap between days 35 and
50 is partly due to a major solar flare.  Below we refer to these two
data sets as \early and \late periods.

\begin{figure}
\begin{center}
\includegraphics[trim= 0cm 5cm 0cm 2cm,
  width=1\textwidth,clip=t,angle=0.,scale=0.49]{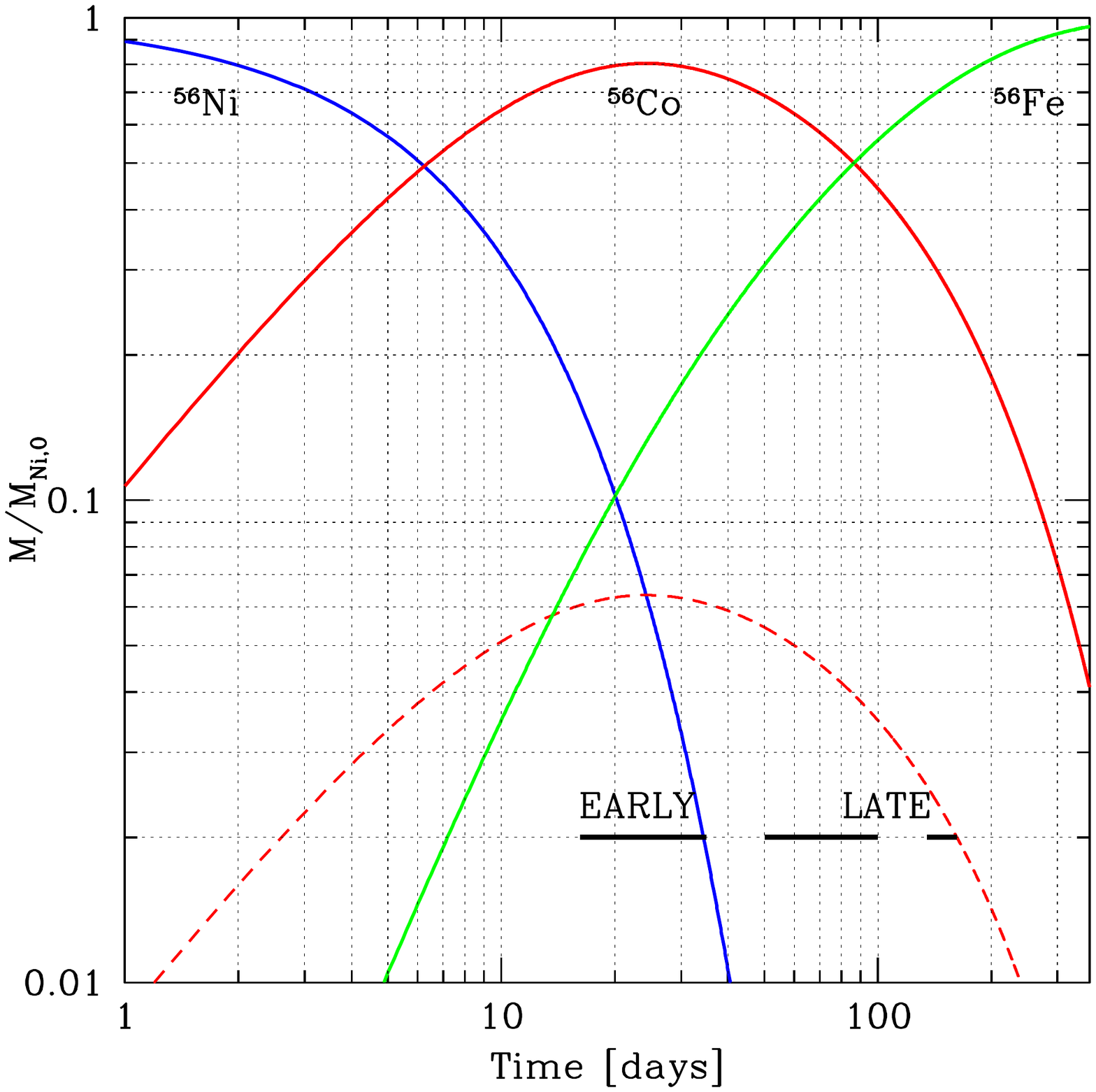}
\end{center}
\caption{\early and \late periods of \INTEGRAL observations
  used for spectra extraction, shown as thick horizontal bars. Three curves show the evolution of the
   $^{56}$Ni, $^{56}$Co and $^{56}$Fe masses, respectively, normalized to the
  initial $^{56}$Ni mass. Note that opacity effects tend to suppress
  the emergence of gamma-rays at early phases of the supernova
  evolution, unless radioactive isotopes are present in the outer
  layers of the ejecta, or the explosion is strongly asymmetric.
  The dashed red line shows the $^{56}$Co mass scaled down by the ratio of
  Co and Ni decay times $\tau_{Co}/\tau{Ni}$, which allows one to
  compare the expected relative strength of Ni (blue curve) and Co
  (dashed red curve) gamma-ray lines as a function of time.
\label{fig:periods}}
\end{figure}

\begin{deluxetable*}{rccc}
\tabletypesize{\footnotesize}
\tablecaption{Data sets}
\tablewidth{0pt}
\tablehead{
\colhead{Set} &
\colhead{Dates} &
\colhead{Days since explosion} &
\colhead{Exposure\tablenotemark{$\alpha$}, Msec}
}
\startdata
\early &   2014-01-31 : 2014-02-20    &  ~16 : ~35 &   1.0\\ 
\late &    2014-03-05  : 2014-06-25 &    ~50 : 162 &  4.3
\enddata
\tablenotetext{$\alpha$}{Corrected for the periods of high background and the dead-time of SPI}
\label{tab:sets}
\end{deluxetable*}

\begin{figure*}
\begin{center}
\includegraphics[trim= 0cm -2cm 0cm 0cm,
  width=1\textwidth,clip=t,angle=0.,scale=0.90]{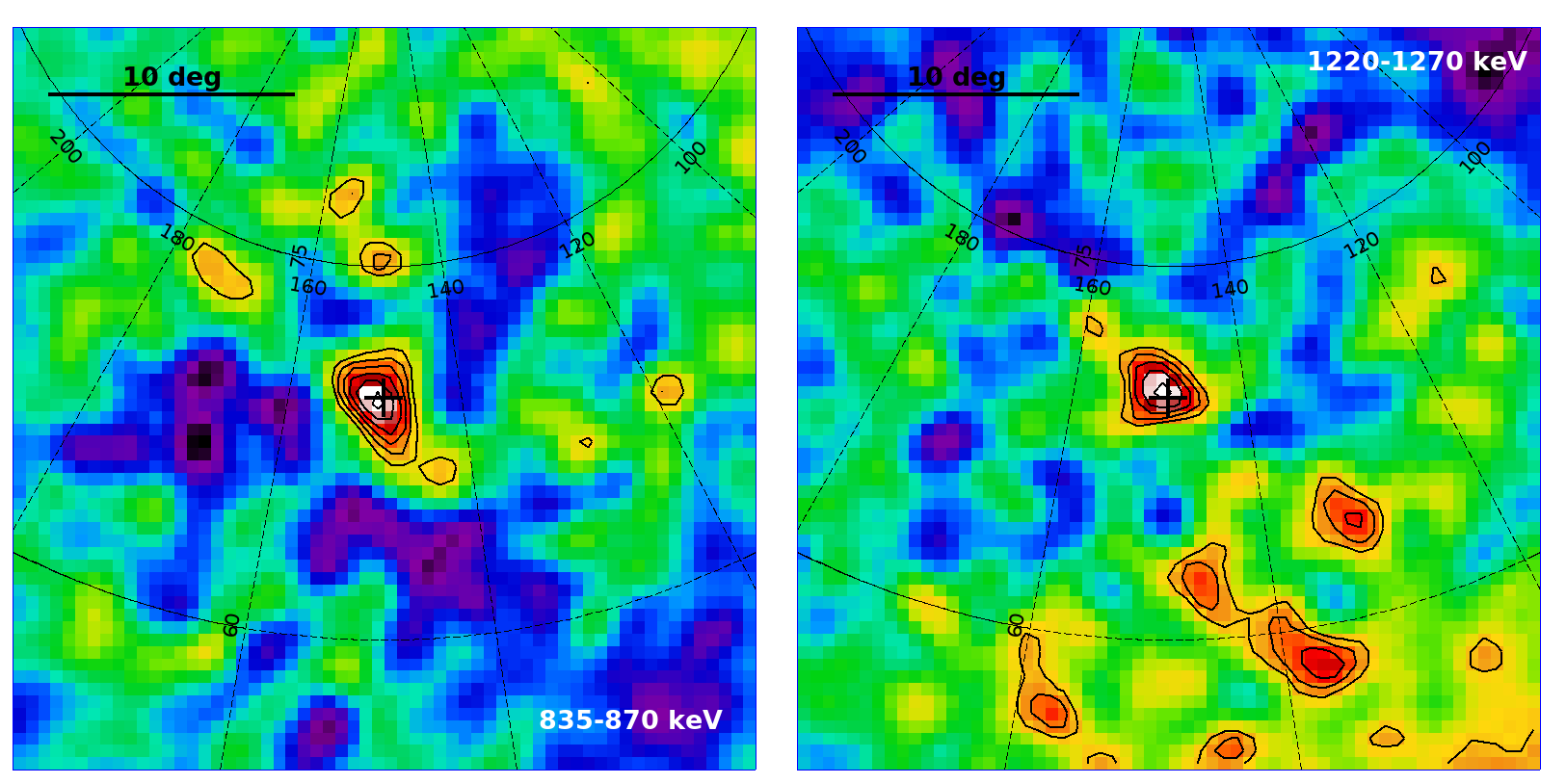}
\end{center}
\caption{SPI images (S/N ratio) during \late period in two narrow
  bands around most prominent $^{56}$Co lines. Contours are at
  2, 2.5 ... 5 $\sigma$. Cross shows the position of SN2014J. The
  brightest peaks in each image coincide well with the position of
  SN2014J. Due to the dither pattern\footnote{http://www.cosmos.esa.int/web/integral} used during observations of SN2014J the central part of the image is much better covered than the outer regions. It is therefore not surprising that the level of noise is increasing away from the nominal target.  
\label{fig:spi_image}}
\end{figure*}

Unlike the \early period, when emergence of the $^{56}$Ni lines
strongly depends on the distribution of the radioactive Ni through the
ejecta, for the \late period the emission in $^{56}$Co lines is a
generic prediction of all plausible models. Two energy bands optimal
for detection of the SN signal in gamma-rays are clear from
Fig.\ref{fig:background}. These two bands, containing the  most
prominent $^{56}$Co lines,  were used to generate images. 
The images were extracted from SPI data from the \late period as in
\citep{2014Natur.512..406C}. Namely, we vary the assumed position of the source and
repeat the flux fitting procedure (see \S\ref{sec:spi}) for each
position. The resulting images of the signal-to-noise ratio in the
835-870 and 1220-1270 keV energy bands are shown in
Fig.\ref{fig:spi_image}. In both energy bands the highest peaks (4.7 and 4.3 $\sigma$
respectively) coincide well (within 0.3\deg) with the
SN2014J position, marked by a cross.

\begin{figure*}
\begin{center}
\includegraphics[trim= 0cm 5cm 0cm 10cm,
  width=1\textwidth,clip=t,angle=0.,scale=0.9]{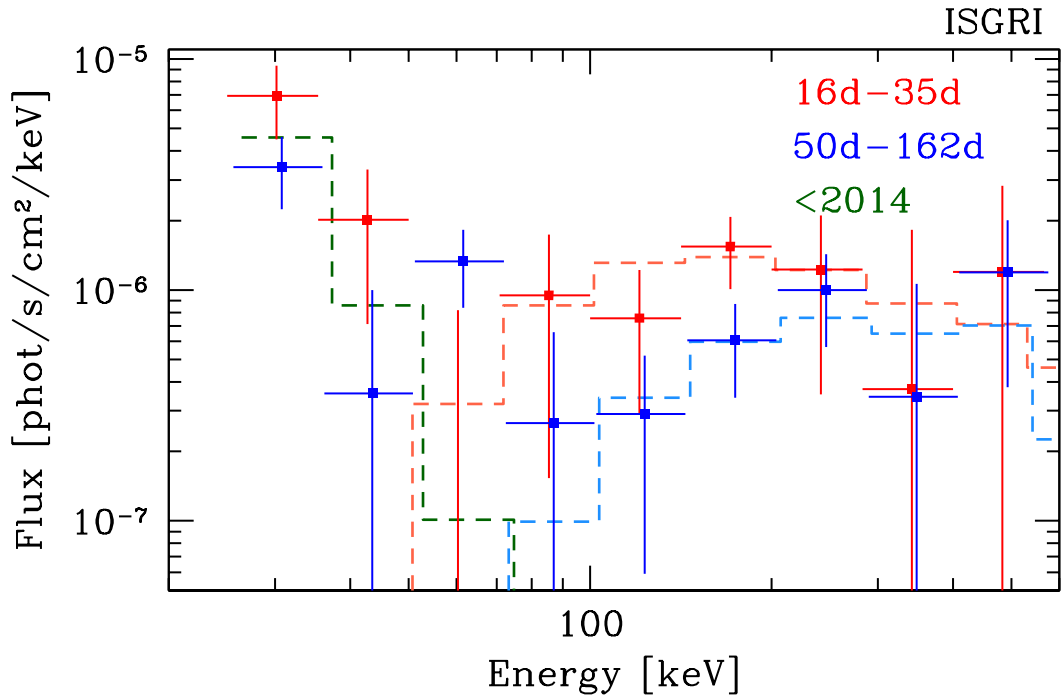}
\end{center}
\caption{ISGRI spectrum measured at the position of SN2014J during
  \early (red) and \late (blue) periods. The energies of the second
  set of points are multiplied by a factor 1.02 for the sake of
  clarity. Dashed histograms show the predicted spectra of the \W7 model for
  the same periods. The agreement with the predictions is reasonable
  except for the energies lower than $\sim 70$ keV, where the spectrum
  is likely contaminated by other sources in M82 \citep[see, e.g,][]{2014AstL...40...65S}. Dark green line shows crude approximation of the M82 spectrum measured before 2014. 
\label{fig:spec_isgri}}
\end{figure*}

The ISGRI spectra extracted at the known position of SN2014J for the
\early and \late period are shown in Fig.~\ref{fig:spec_isgri}. Low
energy (less than $\sim$70 keV) part of the extracted spectrum is
likely contaminated by other sources in M82.

\begin{figure*}
\begin{center}
\includegraphics[trim= 0cm 4cm 0cm 11cm,
  width=1\textwidth,clip=t,angle=0.,scale=0.99]{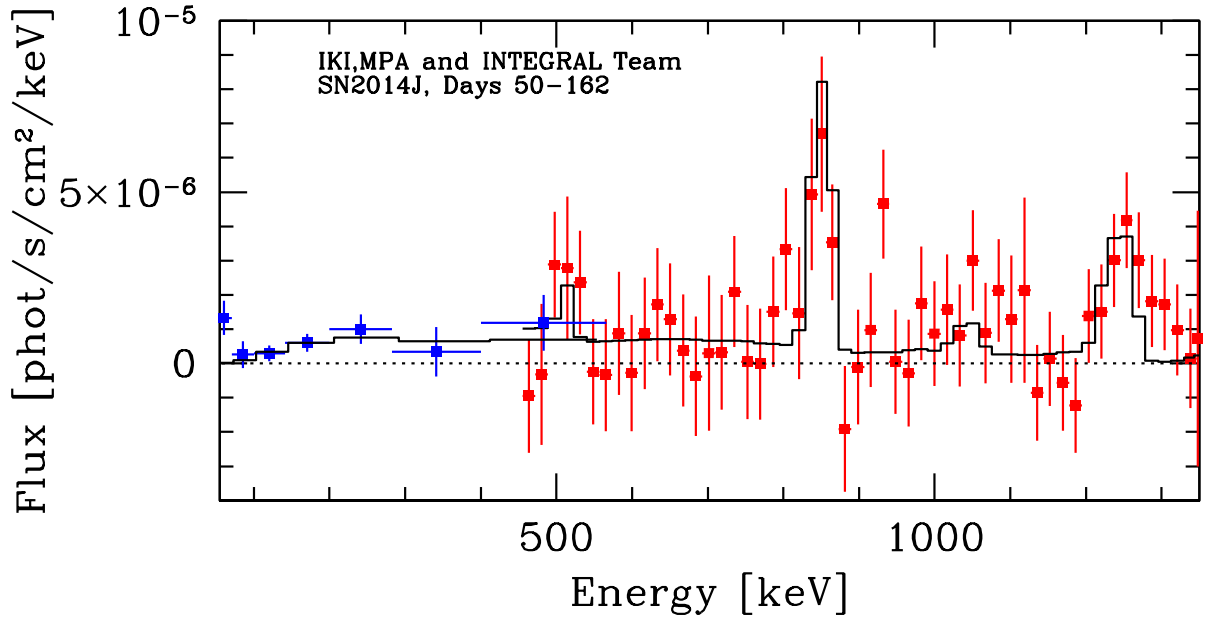}
\end{center}
\caption{Combined ISGRI/SPI spectrum for the \late period. The
  model (\W7, see Tab.\ref{tab:models}) has been convolved with the SPI
  off-diagonal response. The
  SPI data below 450 keV are omitted since during \late period the data at these energies are
  expected to be dominated by the off-diagonal response of SPI.
\label{fig:spec_flate}}
\end{figure*}

\begin{figure*}
\begin{center}
\includegraphics[trim= 0cm 4cm 0cm 11cm,
  width=1\textwidth,clip=t,angle=0.,scale=0.99]{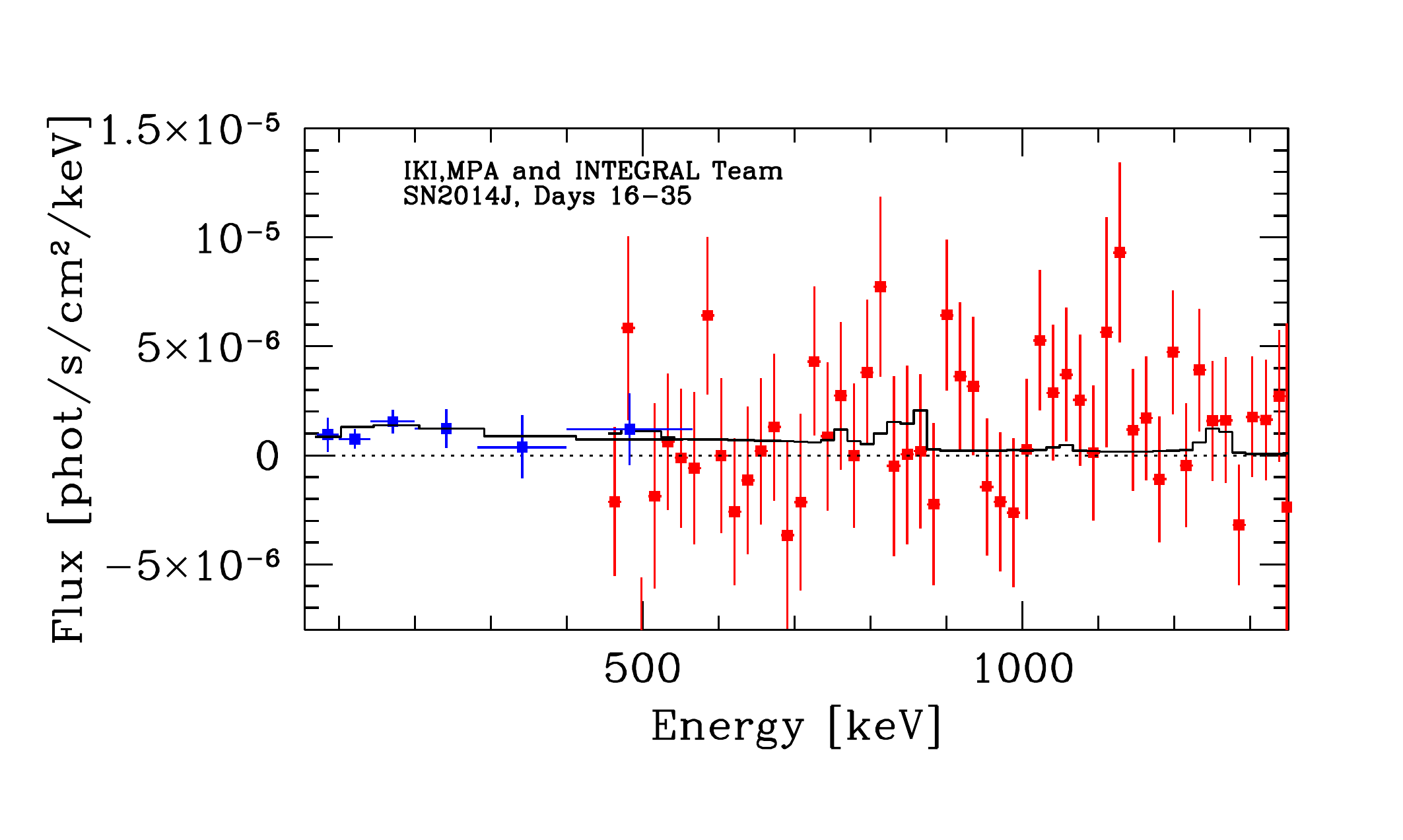}
\end{center}
\caption{Combined ISGRI/SPI spectrum for the \early period. The
  model (\W7, see Tab.\ref{tab:models}) has been convolved with the SPI
  off-diagonal response. 
\label{fig:spec_early}}
\end{figure*}

\section{Models}
\label{sec:models}
\subsection{A set of representative models}
\label{sec:1d}
For comparison with the \INTEGRAL data we used a set of representative
1D models (Table \ref{tab:models}), based on calculations of explosive
nucleosynthesis models. To the first approximation, these models are
characterized by the amount of radioactive nickel, total mass of the
ejecta and the expansion velocity. Although current state-of-the-art
simulations of type Ia explosions can be done in 3D
\citep[e.g.,][]{2013MNRAS.429.1156S,2014MNRAS.438.1762F,2014ApJ...785..105M},
using these models would introduce an additional viewing angle
dependence. In order to avoid this extra degree of freedom and given
that the overall significance of the SN2014J detection in gamma-rays
by \INTEGRAL (see \S\ref{sec:observations} and \S\ref{sec:results})
corresponds to only $\sim 10$ s.t.d., we decided to keep in this work
only a set of 1D models to confront with the data.

\begin{deluxetable}{llll}
\tabletypesize{\footnotesize}
\tablecaption{Set of models used in the paper}
\tablewidth{0pt}
\tablehead{
\colhead{Model} &
\colhead{$M_{Ni},~M_\odot$} &
\colhead{$M_{tot},~M_\odot$} &
\colhead{$E_K,~10^{51}~{\rm erg}$} 
}
\startdata
\texttt{DDT1p1} &     0.54 &      1.36 &       1.29 \\ 
\texttt{DDT1p4halo} & 0.62 &      1.55 &       1.3  \\ 
\DDTe &       0.51 &      1.37 &      1.09\\ 
\DETO &       1.16 &      1.38 &       1.44 \\ 
\HED6 &       0.26 &      0.77 &       0.72 \\ 
\W7 &        0.59 &      1.38 &       1.24 \\ 
\texttt{ddt1p4} &     0.66 &      1.36 &       1.35 \\ 
\BALL3 &     0.66+0.04\tablenotemark{$\alpha$}  &      1.36 &       1.35\\ 
\DD4 &        0.61 &      1.39 &       1.24   
\enddata
\tablenotetext{$\alpha$}{additional ``plume'' of $^{56}$Ni.}
\label{tab:models}
\end{deluxetable}

The set of models includes the deflagration model \W7
\citep{1984ApJ...286..644N}, pure detonation model \DETO
\citep{2003ApJ...593..358B}, the sub-Chandrasekhar model \HED6
\citep{1996ApJ...457..500H}, and several variants of the delayed
detonation models: \DD4 \citep{ww91}, \DDTe
~\citep{2003ApJ...593..358B}, \texttt{DDT1p1}, \texttt{DDT1p4halo},
\texttt{ddt1p4}, \BALL3 \citep{isern}. The \texttt{ddt1p4} model was
built to match the mass of $^{56}$Ni suggested by the early optical
evolution of SN2014J 
as detected with the OMC of INTEGRAL (\citet{2014ATel.6099....1I};
 P. Hofflich, private communication). In it, the
transition density from deflagration to detonation was fixed at
$1.4~10^7~{\rm g~cm^{-3}}$. Model \texttt{DDT1p4halo} is a variant of
the later in which the white dwarf is surrounded by a 0.2 $M_\odot$
envelope, as might result from a delayed merger explosion. The \BALL3
model is essentially the same as the \texttt{ddt1p4} plus a plume of
$0.04~M_\odot$ of radioactive $^{56}Ni$ receding from the observer
\citep[see][for details]{isern}.

The emerging X-ray and gamma-ray radiation from the expanding SNIa is
determined by the total amount of radioactive isotopes, their
distribution over velocities, the mass and the chemical composition of
the ejecta and expansion rate. The processes are essentially the same
as in type II supernovae \citep[see, e.g.][ for a prototypical example
  of type II supernova - SN1987A]{1987Natur.330..227S}. However, the
mass of the ejecta and expansion rate differ strongly leading to much
earlier and stronger signal in gamma-rays \citep[see,
  e.g.][]{1969ApJ...155...75C,1981CNPPh...9..185W,1988ApJ...325..820A}. A
comprehensive set of computations of the expected gamma-ray flux for
different representative models was presented in \citet{2014ApJ...786..141T}.

Here we use the
results of similar calculations (see below), which account for line broadening,
needed for systematic comparison with the \INTEGRAL data. 

A Monte-Carlo code follows the propagation of the $\gamma-$photons
through the ejecta and accounts for scattering and photoabsorption of
photons and annihilation of positrons. The predicted spectra were
generated with a time step of one day, covering the entire
observational period. These model spectra were then averaged over the
periods of 16-35 and 50-162 days respectively, to provide fair
comparison with the \INTEGRAL results for the \early and \late
periods. In particular, the effect of varying opacity in each model over the observational period is correctly captured by this procedure.  
The computations include full treatment of Compton scattering
(coherent and incoherent), photoabsorption and pair production
\citep[see][for details]{2004ApJ...613.1101M}. The positrons produced by $\beta^+$
decay of $^{56}$Co (19\% of all decays) annihilate in place via
positronium formation. Both two-photon annihilation into the 511 keV
line and the orthopositronium continuum are included.

\subsection{Transparent ejecta model (\tem)}
\label{sec:tem}
As we discuss below (\S\ref{sec:results}) the \INTEGRAL data are
broadly consistent with the subset of models listed in
Table~\ref{tab:models}. However, \citet{2014Sci...345.1162D} reported
an evidence of $^{56}$Ni at the surface in the first observations of
SN2014J with \INTEGRAL \citep[see also][for an alternative analysis of
  early SN2014J observations]{isern}. Presence of radioactive material
at the surface would be an important result, since traditional models,
listed in Table~\ref{tab:models} do not predict it. One can
  attempt to patch our 1D models with an additional component
  describing an extra radioactive material at the surface. Assuming
  that the material at the surface is transparent to gamma-rays, the
  fluxes of individual lines associated with Ni and Co decay, their
  energies and widths can be tied together. The transparency
  assumption is justified by the large velocities and small initial
  densities expected for matter at the surface of supernovae
  ejecta. In any case, it provides a lower limit to the mass of
  radioactive material, as opacity would demand a larger gamma-ray
  production rate in order to explain a given gamma-ray flux.
  This approach allows to describe many
  lines, associated with a transparent clump with only 3
  parameters. Below we refere to this component as a Transparent
  Ejecta model (\texttt{TEM}), and use it in combination with the
  best-performing \W7 model from out default set 1D models (see
  \S\ref{sec:1d}), i.e., the data are compared with the predictions of
  \W7+\texttt{TEM} model.  While this model by itself is not
  selfconsistent, it can be used to answer the following questions:
  \begin{itemize}
  \item Once the predicted signal for the \W7 model is removed from
    the observed spectra, do residuals resemble a signal expected from
    a transparent clump of radioactive material?
  \item Given the statistics accumulated by {\it INTEGRAL}, how much
    radioactive material in a transparent clump can be ``hidden'' in
    the data on top of a given 1D model?
  \end{itemize}
In this section we describe the \texttt{TEM} model and then apply it
to the data in \S\ref{sec:tem_late}.

The \texttt{TEM} model assumes that all line energies are shifted
proportionally to their energies (i.e., the same velocity structure
for all lines), while their flux ratios follow the predicted ratios
\citep{1994ApJS...92..527N} based on the decay chains of
$^{56}$Ni$\rightarrow^{56}$Co$\rightarrow^{56}$Fe and
$^{57}$Ni$\rightarrow^{57}$Co$\rightarrow^{57}$Fe. The list of the
lines and their fluxes normalized to 1~$M_\odot$ of $^{56}$Ni are
given in Table~\ref{tab:tem}. For a given time period the model has 3
parameters: the initial $^{56}$Ni mass ($M_{Ni}$), energy/redshift of
the 847 keV line ($E_{847}$) and the broadening of the 847 keV line
($\sigma_{847}$).  The width of each line (Gaussian $\sigma$) is
defined as
\begin{eqnarray}
\sigma_{line}=\sigma_{847}\times \left(
  \frac{E_{line}}{E_{847}}\right ).
\end{eqnarray}
Ortho-positronium continuum and pair
production by gamma-ray photons are
neglected, while the 511 keV line is added assuming that 19\% of
$^{56}$Co decays produce positrons, of which 25\% form
para-positronium yielding two 511 keV photons.

\begin{deluxetable}{rlll}
\tabletypesize{\footnotesize}
\tablecaption{Line fluxes averaged over days 50-162 for a transparent
  ejecta model (\tem) for the initial $1~M_\odot$ of $^{56}$Ni}
\tablewidth{0pt}
\tablehead{
\colhead{$E_{line}$, keV} &
\colhead{$F_{line}/F_{847}$} &
\colhead{Flux$^a$} &
\colhead{Isotope} 
}
\startdata
    846.78 & $  1.00 $ & $  6.57~10^{-4}$ & $^{56}$Co \\ 
    158.38 & $  7.98~10^{-3}$ & $  5.25~10^{-6}$ & $^{56}$Ni \\ 
   1561.80 & $  1.12~10^{-3}$ & $  7.34~10^{-7}$ & $^{56}$Ni \\ 
    749.95 & $  3.99~10^{-3}$ & $  2.62~10^{-6}$ & $^{56}$Ni \\ 
    269.50 & $  2.87~10^{-3}$ & $  1.89~10^{-6}$ & $^{56}$Ni \\ 
    480.44 & $  2.87~10^{-3}$ & $  1.89~10^{-6}$ & $^{56}$Ni \\ 
    811.85 & $  6.86~10^{-3}$ & $  4.51~10^{-6}$ & $^{56}$Ni \\ 
    511.00 & $  9.50~10^{-2}$ & $  6.24~10^{-5}$ & $^{56}$Co \\ 
   1037.83 & $  1.40~10^{-1}$ & $  9.20~10^{-5}$ & $^{56}$Co \\ 
   1238.28 & $  6.80~10^{-1}$ & $  4.47~10^{-4}$ & $^{56}$Co \\ 
   $^*$1771.49 & $  1.60~10^{-1}$ & $  1.05~10^{-4}$ & $^{56}$Co \\ 
   $^*$2034.92 & $  7.90~10^{-2}$ & $  5.19~10^{-5}$ & $^{56}$Co \\ 
   $^*$2598.58 & $  1.69~10^{-1}$ & $  1.11~10^{-4}$ & $^{56}$Co \\ 
   $^*$3253.60 & $  7.40~10^{-2}$ & $  4.86~10^{-5}$ & $^{56}$Co \\ 
   $^*$14.41 & $  1.19~10^{-3}$ & $  7.80~10^{-7}$ & $^{57}$Co \\ 
    122.06 & $  1.03~10^{-2}$ & $  6.79~10^{-6}$ & $^{57}$Co \\ 
    136.47 & $  1.19~10^{-3}$ & $  7.80~10^{-7}$ & $^{57}$Co
\tablecomments{$^a$ - Flux is in units of $~{\rm phot~s^{-1}~cm^{-2}}$ \\
$^*$ - Line is outside the energy range used for fitting} \\
\label{tab:tem}
\end{deluxetable}

\begin{figure}
\begin{center}
\includegraphics[trim= 0cm 5cm 0cm 2cm,
  width=1\textwidth,clip=t,angle=0.,scale=0.49]{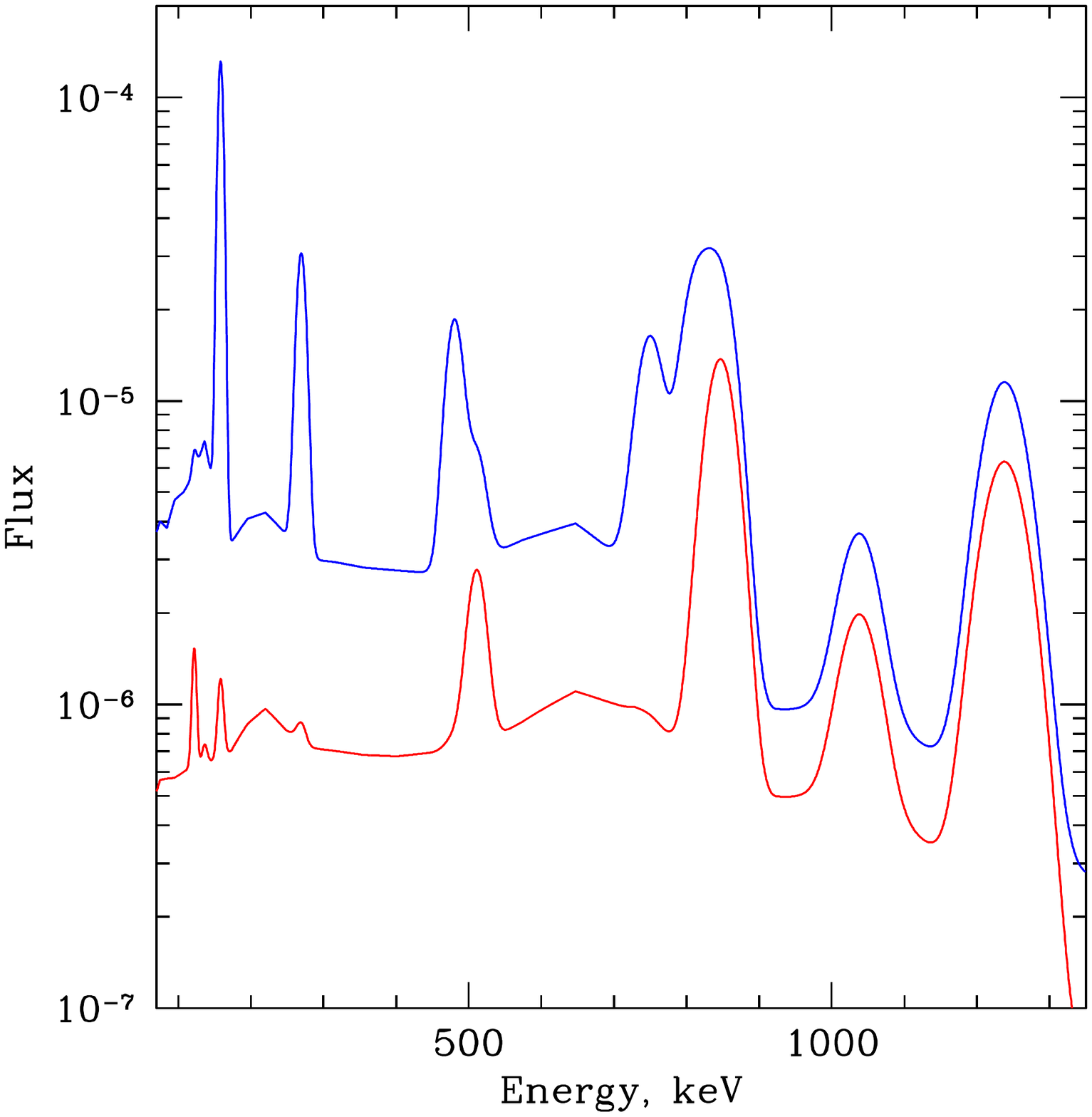}
\end{center}
\caption{Spectra predicted by the \tem model for \early (blue) and
  \late (red) data sets, convolved with the SPI response. The
  broadening of the reference 847 keV line is set to 20 keV (Gaussian
  sigma). The initial $^{56}$Ni mass is 1 $M_\odot$.
\label{fig:tem}}
\end{figure}

\section{Results}
\label{sec:results}
\subsection{Combined ISGRI+SPI spectrum}
\label{sec:spectra}
The SPI images (Fig.\ref{fig:spi_image}) for \late period unambiguously
show the characteristic signatures of $^{56}$Co decay from SN2014J. A
more quantitative statement on the amount of $^{56}$Ni synthesized
during explosion and on the properties of the ejecta can be obtained
from the comparison of the data with the predictions of the
models. Since the \late period is less affected by the transparency of
the ejecta we start our analysis with the total spectrum obtained by
\INTEGRAL over this period.

\subsubsection{{\bf Late} data}
The results of fitting of the combined ISGRI+SPI spectrum
(Fig.\ref{fig:spec_flate}) for the \late period are given in Table
\ref{tab:mfit_late}. A full set of
models from Tab.\ref{tab:models} is used. The two groups of columns in Table~\ref{tab:mfit_late} differ by the energy range in the SPI data used for comparison with the model. In the first group  the data of ISGRI (70-600 keV) and SPI (400-1350 keV) are used.  The
data below 70 keV are likely contaminated by other sources in M82. The
SPI data below 400 keV are omitted since during the \late period the data at
these energies are expected to be dominated by the off-diagonal
response of SPI. I.e. the observed SPI spectrum below 500 keV includes
significant contribution of the gamma-ray photons at higher energies,
which are down-scattered inside the body of the telescope (see
Fig.\ref{fig:offdiag}). The Null model (no source) gives $\chi^2=1945.38$ for 1906 spectral
  bins. The improvement of the $\chi^2$ relative to the Null is
  calculated by fixing the normalization at the predicted value for
  $D=3.5$ Mpc (column 2) and by letting it free (columns 3 and
  4). The typical value of the $\Delta \chi^2\sim 65$ suggests $\sim
  8~\sigma$ detection. 

One can draw two conclusions from this exercise. First of all a set of
canonical 1D deflagration (\W7) or delayed detonation models (e.g., \DD4)
fit the data well without any adjustments to the normalization. The
pure detonation model \DETO and a sub-Chandrasekhar model \HED6 give
poor fit and overproduce/underproduce the observed flux,
respectively. Secondly, once the normalization is allowed to vary, all
models give almost identical gain in the $\chi^2$, suggesting that
relative strength of all prominent features is comparable in all
models. Given the uncertainty in the distance to SN2014J (or M82) a
deviation of the normalization at the level of $\sim$20\% can not be
excluded. But \DETO and \HED6 models require by far larger changes in
the normalization.

\begin{deluxetable*}{lcrrcrr}
\tabletypesize{\footnotesize}
\tablecaption{$\Delta\chi^2$ for basic models for fixed and free
  normalization relative to the Null model of no source for the \late period. }
\tablewidth{12cm}
\tablehead{
\colhead{Dataset:} & \multicolumn{3}{c}{ISGRI(70-600 keV)\&SPI(400-1350 keV)}    &        \multicolumn{3}{c}{ISGRI(70-600 keV)\&SPI(70-1350 keV)}\\
\\  
\colhead{Model} &
\colhead{$N=1,\Delta\chi^2$} &
\colhead{$N_{free}$} &
\colhead{$\Delta\chi^2$} &
\colhead{$N=1,\Delta\chi^2$} &
\colhead{$N_{free}$} &
\colhead{$\Delta\chi^2$}
}
\startdata
\texttt{DDT1p1} &      66.4 &       1.03$\pm$      0.13 &      66.5  &    87.3 &       1.09$\pm$      0.12 &      87.9  \\      
\texttt{DDT1p4halo} &  65.9 &       0.89$\pm$      0.11 &      66.9  &    88.1 &       0.93$\pm$      0.10 &      88.5  \\  
\DDTe &                62.1 &       1.09$\pm$      0.14 &      62.5  &    82.3 &       1.15$\pm$      0.13 &      83.7  \\        
\DETO &                10.1 &       0.52$\pm$      0.06 &      66.4  &    30.2 &       0.55$\pm$      0.06 &      87.7  \\        
\HED6 &                47.8 &       1.86$\pm$      0.24 &      60.7  &    60.1 &       2.01$\pm$      0.22 &      80.5  \\        
\W7 &                  65.0 &       0.94$\pm$      0.12 &      65.3  &    86.9 &       1.01$\pm$      0.11 &      86.9  \\          
\texttt{ddt1p4} &      64.9 &       0.85$\pm$      0.10 &      66.9  &    87.4 &       0.90$\pm$      0.10 &      88.4  \\      
\BALL3 &               63.2 &       0.83$\pm$      0.10 &      66.1  &    85.7 &       0.88$\pm$      0.09 &      87.5  \\     
\DD4 &                 64.7 &       0.89$\pm$      0.11 &      65.7  &    87.0 &       0.95$\pm$      0.10 &      87.3 \\
               &           &                           &            &         &                           &           \\ 
\texttt{No source}, $\chi^2$ (d.o.f.) &            & 1945.4 (1906)             &       &    & 2696.9 (2566)         &       
\enddata

\tablecomments{$N$ is the normalization of the model with $N=1$ corresponding to the explosion at the distance of 3.5 Mpc.
  $\Delta\chi^2$ characterizes an improvement of $\chi^2$
  for a given model relative to the Null model. Larger positive values
  indicate that the model is describing the data significantly better than other models (see Appendix).
  The data below 70 keV are likely contaminated by other sources
  in M82. SPI data below 400 keV 
  are omitted in the first dataset (left half of the Table) since the data at these energies are
  expected to be dominated by the off-diagonal response of SPI (see \S\ref{sec:spi}).}
\label{tab:mfit_late}
\end{deluxetable*}

While in the above analysis the SPI data with $E<400$ keV have been omitted to
concentrate on the data less affected by the off-diagonal response, the right part of Tab.\ref{tab:mfit_late} extends the analysis down to 70 keV for
both instruments. The basic conclusions remain the same, although, as
expected, the significance of the detection increases to $\gtrsim
9~\sigma$. 

\subsubsection{{\bf Early} data}
We now proceed with the same analysis of the \early data. Table
\ref{tab:mfit_early} contains the gain in the $\chi^2$ for the same
set of models.

The \DETO is clearly inconsistent with the data - inclusion of the
model increases the $\chi^2$ relative to the Null model (no source).
The \HED6 model, which gave a poor fit to the \late data, yields the
$\chi^2$ comparable to other models. This is because smaller amount of
$^{56}$Ni is compensated by larger transparency of the lower-mass
ejecta, which is important for the \early data.

The \BALL3 model gives poor gain in $\chi^2$ if the normalization is
fixed and the SPI data below 400 keV are excluded. If the
normalization is free, and, especially, if the SPI data below 400 keV
are included, this model performs marginally better than other
models. However, it performs significantly better than other models
  when the SPI data below 400 keV are included. This is not surprising,
  since \BALL3 model has been designed to fit the SPI data during this
  period \citep[see][for details]{isern}. The different ``ranking'' of
  the \BALL3 model seen in Table \ref{tab:mfit_early} when SPI data
  below 400 keV are included or excluded, suggests a tension in the
  comparison of the fixed-normalization \BALL3 model with the SPI and
  ISGRI data and also with the SPI data below and above 400 keV (see below).

\begin{deluxetable*}{lcrrcrr}
\tabletypesize{\footnotesize}
\tablecaption{The same as in Table \ref{tab:mfit_late} for the \early period.}
\tablewidth{12cm}
\tablehead{
\colhead{Dataset:} & \multicolumn{3}{c}{ISGRI(70-600 keV)\&SPI(400-1350 keV)}    &        \multicolumn{3}{c}{ISGRI(70-600 keV)\&SPI(70-1350 keV)}\\
\\  
\colhead{Model} &
\colhead{$N=1,\Delta\chi^2$} &
\colhead{$N_{free}$} &
\colhead{$\Delta\chi^2$} &
\colhead{$N=1,\Delta\chi^2$} &
\colhead{$N_{free}$} &
\colhead{$\Delta\chi^2$}
}
\startdata
\texttt{DDT1p1} &             14.9 &       0.84$\pm$      0.21 &      15.4  &             33.2 &       1.11$\pm$      0.19 &      33.5  \\    
\texttt{DDT1p4halo} &         14.6 &       1.00$\pm$      0.26 &      14.6  &             29.8 &       1.34$\pm$      0.24 &      31.8  \\ 
\DDTe &               14.3 &       1.30$\pm$      0.33 &      15.1  &             26.9 &       1.72$\pm$      0.30 &      32.6  \\       
\DETO &              -83.9 &       0.28$\pm$      0.07 &      14.8  &            -64.8 &       0.37$\pm$      0.06 &      35.2  \\       
\HED6 &               15.7 &       1.05$\pm$      0.26 &      15.8  &             32.7 &       1.39$\pm$      0.23 &      35.5  \\       
\W7 &                 15.9 &       0.87$\pm$      0.22 &      16.2  &             34.8 &       1.14$\pm$      0.19 &      35.3  \\         
\texttt{ddt1p4} &             11.3 &       0.65$\pm$      0.17 &      15.7  &             33.3 &       0.86$\pm$      0.15 &      34.2  \\     
\texttt{3Dbball} &             6.7 &       0.56$\pm$      0.13 &      17.6  &             37.0 &       0.76$\pm$      0.12 &      41.4  \\    
\DD4 &                14.1 &       0.77$\pm$      0.19 &      15.5  &             33.6 &       1.01$\pm$      0.17 &      33.6  \\          
&           &                           &            &         &                           &           \\ 
\texttt{No source}, $\chi^2$ (d.o.f.) &            & 1856.7 (1906)             &       &    & 2615.9 (2566)         &       
\enddata
\label{tab:mfit_early}
\end{deluxetable*}

\begin{deluxetable*}{lcc}
\tabletypesize{\footnotesize}
\tablecaption{$\Delta\chi^2$ for the joint data set of the  \early and \late spectra for a basic set of models with fixed normalization. The value of $\Delta\chi^2$ shows the improvement of the $\chi^2$  relative to the Null model of no source. 
}
\tablewidth{12cm}
\tablehead{
\colhead{Model} &
\colhead{ISGRI \& SPI(400-1350 keV)} &
\colhead{ISGRI \& SPI(70-1350 keV)}\\
\colhead{} &
\colhead{$\Delta\chi^2$} &
\colhead{$\Delta\chi^2$} 
}
\startdata
\texttt{DDT1p1} &      \bf{81.3} &     \bf{120.5} \\ 
\texttt{DDT1p4halo} &      \bf{80.5} &     117.8 \\ 
\DDTe &      76.4 &     109.2 \\ 
\DETO &     -73.8 &     -34.7 \\ 
\HED6 &      63.5 &      92.8 \\ 
\W7 &      \bf{80.9} &     \bf{121.7} \\ 
\texttt{ddt1p4} &      76.2 &     \bf{120.7} \\ 
\texttt{3Dbball} &      69.9 &     \bf{122.7} \\ 
\DD4 &      \bf{78.8} &     \bf{120.7}  
\enddata
\tablecomments{Bold-faced are the models which have $\Delta\chi^2$ different from the model with the largest $\Delta\chi^2$ by less than 4, the criterion used to group models into ``more plausible'' and ``less plausible'' respectively (see Appendix).}
\label{tab:mfit_eandl}
\end{deluxetable*}

\subsubsection{{\bf Early} and {\bf Late} data together}
Finally, in Table~\ref{tab:mfit_eandl} we compare jointly the \early
and \late data of ISGRI and SPI with the models, calculated for
corresponding periods. The two columns in Table~\ref{tab:mfit_eandl}
differ by the energy range in the SPI data used for comparison with
the model. In each case the normalization was fixed at the value set
by the adopted distance of 3.5 Mpc. In each column we mark with bold
face the models which have $\Delta\chi^2$ different from the model
with the largest $\Delta\chi^2$ by less than 4 (see Appendix for the clarification on the interpretation of this criterion in Bayesian and frequentist approaches). Once again, 1D deflagration model \W7 and
"standard" delayed detonation model perform well. The \BALL3, which
was designed to account for tentative feature in the \early SPI data
at low energies, not surprisingly performs well if the SPI data below
400 keV are included. However, if only the data above 400 keV are used
for SPI, this model yields significantly lower $\Delta\chi^2$ than the
\W7 or \texttt{DDT1p1} models.

\subsection{Comparison of gamma-ray light curves with models}
\label{sec:lightcurves}
While the spectra for the \early and \late periods already provide an
overall test of the basic models, additional information can be
obtained by analyzing the time variations of the fluxes in broad
energy bands (see Fig.\ref{fig:lc_isgri} and \ref{fig:lc_spi}). The
total number of time bins is 34.  Each bin corresponds to one
revolution (i.e. $\sim$3 days). The first raw in
Table~\ref{tab:lc_chi2} provides the values of the $\chi^2$ (for Null
model of no source) in three energy bands: 100-200 keV (ISGRI),
835-870 keV (SPI) and 1220-1272 keV (SPI). The normalization of the
model lightcurves is fixed to 1.  For 34 bins the value $\chi^2$ for a
correct model is expected to be in the interval $\sim$26-42 in 68\% of
cases. Clearly, the Null model does not fit the data well.

Other raws show the improvement of the $\chi^2$ relative to the Null
model. I.e., $\displaystyle \Delta \chi^2=\chi^2_{Null}-\chi^2_{model}$.  From
Table~\ref{tab:lc_chi2} it is clear that \DETO model strongly
overpredicts the flux in all bands and can be excluded ($\chi^2$
becomes worse when this model is used). Other models leads to significant
improvement with respect to the Null model, except for the \BALL3
model in the 100-200 keV band where it exceeds the observed flux in
the early observation, while in the SPI bands all these models are
comparable.

The last column in Table~\ref{tab:lc_chi2} provides the $\chi^2$ for
three bands joints. This is basically the sum of the values of
$\chi^2$ for individual bands. Bold-faced are the best performing
models: \W7 and \texttt{DDT1P1}. As in \S\ref{sec:spectra} these are
the models which have $\Delta\chi^2$ different from the model with the
largest $\Delta\chi^2$ by less than 4 (see Appendix).

One can also compare the lightcurves with the hypothesis of a
  constant flux. The mean level of flux was estimated for each band
  and the value of the $\chi^2$ was calculated. The values of
  $\Delta\chi^2$ relative to ``No source'' are given in the last row
  of Table~\ref{tab:lc_chi2}. One can see that this simple model is
  almost as good as other best-performing models in individual bands
  (even taking into account that this model has a free parameter -
  mean flux).  This is of course the result of low statistical
  significance of the SN2014J detection that makes it difficult to
  constrain time variations of a faint signal. For the combined values
  for all three band the effective number of free parameter is 3 (mean
  fluxes in each band) and one can conclude that, e.g. \W7 model
  performs marginally better than the constant flux model.

\begin{deluxetable*}{lrrrr}
\tabletypesize{\footnotesize}
\tablecaption{$\Delta\chi^2$ for light-curves in three energy bands for
  different models. The value of $\Delta\chi^2$ shows the improvement of the $\chi^2$  relative to the Null model of no source. The value of the  $\chi^2$ for the Null model is given in the first raw.
}
%\tablewidth{12cm}
\tablehead{
\colhead{Model} &
\colhead{100-200 keV (ISGRI)} &
\colhead{835-870 keV (SPI)} &
\colhead{1220-1272 keV (SPI)} &
\colhead{Three bands jointly} 
}
\startdata
\texttt{No source}, $\chi^2$ &      51.0 &      49.6 &      51.1 &     151.7\\ 
\\
\texttt{DDT1P1} &      16.9 &      18.6 &      19.3 &      {\bf 54.8}\\ 
\texttt{DDT1P4halo} &       9.2 &      17.4 &      19.5 &      46.1\\
\DDTe &      16.7 &      17.1 &      17.6 &      51.4\\ 
\DETO &    -105.0 &      -6.3 &      13.5 &     -97.8\\ 
\HED6 &      20.6 &      16.0 &      14.1 &      50.7\\ 
\W7 &      18.8 &      18.4 &      20.0 &      {\bf 57.2}\\ 
\texttt{DDT1P4} &       9.1 &      17.5 &      20.3 &      46.9\\ 
\texttt{3Dbball} &      -4.8 &      17.4 &      20.4 &      33.0\\ 
\DD4 &      12.7 &      17.9 &      20.0 &      50.6 \\
\texttt{CONST} &      18.7 &      16.1 &      18.5 &      53.3
\enddata
\tablecomments{Total number of time bins is 34. For the joint $\chi^2$ the effective number of bins is three times larger - 102. }
\label{tab:lc_chi2}
\end{deluxetable*}

\subsection{Search for the velocity substructure in the \late data}
\label{sec:tem_late}
\label{sec:tem_early}
The above analysis suggests that the \INTEGRAL data broadly agree with
a subset of simple 1D models (e.g., \W7 or \DD4). Since the true
structure of SN2014J is surely more complicated than predicted by 1D
models, it is interesting to verify if adding an extra component to
the model (on top of the best-performing \W7 model) significantly
improves the fit. In this section we use \tem model as such extra
component. This choice is partly driven by the discussion of a
possible presence of $^{56}$Ni at or near the surface of the ejecta in
\citet{2014Sci...345.1162D} and \citet{isern}.  As described in
\S\ref{sec:tem} the \tem model described a transparent clump of
radioactive Ni. All gamma-ray lines associated with the
Ni$\rightarrow$Co$\rightarrow$Fe decay in the \tem model are tied to
the energy (redshift) and the width of the reference 847 keV line.
The flux ratios are also tied together using a model of an optically
thin clump, taking into account time evolution of the Ni and Co
masses. Examples of spectra predicted by \tem model (for 1~$M_\odot$
of $^{56}$Ni) are shown in Fig.~\ref{fig:tem}.

\begin{figure}
\begin{center}
\includegraphics[trim= 0cm 5cm 0cm 2cm,
  width=1\textwidth,clip=t,angle=0.,scale=0.49]{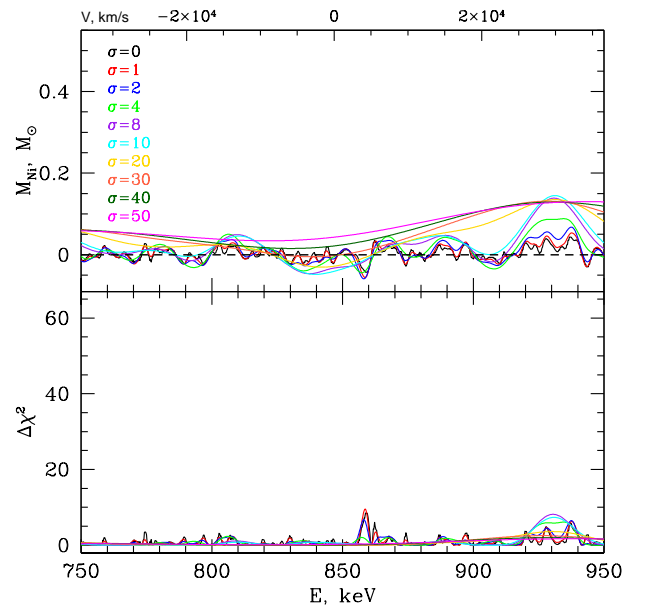}
\end{center}
\caption{ Fitting the SPI data in the 400-1350 keV band with a
    composite \W7 + \tem model. The normalization of the \W7 model is
    fixed to 1. In the \tem model all lines are tied to the energy
    (redshift) and the width of the reference 847 keV line. The flux
    ratios are tied using a model of optically thin clump, taking into
    account time evolution of the Ni and Co masses. This setup is
    optimized for a search of a transparent clump on top of the \W7
    model).  For a given energy and width the model has only
    normalization (initial $^{56}$Ni mass in the clump) as a free
    parameter. The bottom panel shows the improvement in $\chi^2$ and
    the top panel shows the best-fitting $^{56}$Ni clump
    mass. Different colors correspond to a different 847 keV line
    broadening parameterized through a Gaussian $\sigma$ - see legend.
    No compelling evidence for a clump is seen in the data. The
    sensitivity of the data to the mass of the clump strongly depends
    on the broadening of the lines (see Fig.\ref{fig:sigma_m}).
\label{fig:tem_late}}
\end{figure}

Thus, we consider a composite model, consisting from the \W7 model (with
the normalization fixed to 1) and the \tem model. This two-component (\W7+\tem)
model effectively searches for a transparent clump of radioactive
material on top of the base-line \W7 model (see
Fig.\ref{fig:tem_late}). The horizontal axis shows the energy of the
reference 847 keV line in the observer frame and different colors
correspond to different 847 keV line broadening parameterized through
a Gaussian $\sigma$ - see legend. For a given redshift/energy and
width of the reference 847 keV line the model has the normalization
(initial $^{56}$Ni mass) as the only free parameter. The best-fitting
$^{56}$Ni mass is shown in the top panel of
Fig.\ref{fig:tem_late}. The bottom panel (Fig.\ref{fig:tem_late})
shows the improvement in the $\chi^2$ (relative to the \W7 model
alone) due to the \tem model.

\begin{figure}
\begin{center}
\includegraphics[trim= 0cm 5cm 0cm 2cm,
  width=1\textwidth,clip=t,angle=0.,scale=0.49]{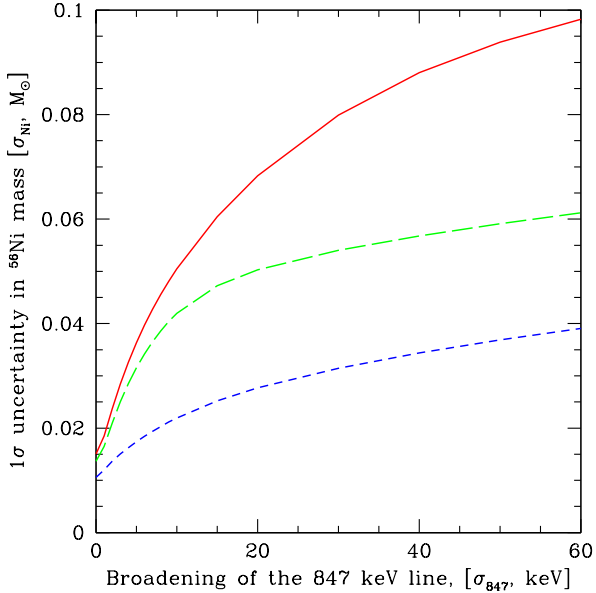}
\end{center}
\caption{Uncertainty in the initial $^{56}$Ni mass as a function of
  line broadening for \early set (dashed-blue: SPI data in the
  70-1350 keV band; long-dashed-green: 400-1350 keV) and \late
  set (solid-red: 400-1350 keV), respectively, assuming transparency to gamma-rays generated close to the surface. A conservative upper limit on the initial mass of "extra" radioactive   $^{56}$Ni, is three times this value at a given line width.  For the line broadening of $10^4~{\rm km~s^{-1}}$ (FWHM), the expected value of $\sigma_{847}$ is $\sim 12~$keV. This value can be regarded as a fiducial value for a simple SNIa model.
\label{fig:sigma_m}}
\end{figure}

As is clear from Fig.~\ref{fig:tem_late} this model does no provide
compelling evidence for a transparent clump on top of the \W7 in the
\late data. Formally, there is a $\Delta \chi^2\sim 9.5$ peak at
$\sim$858.5 keV, which corresponds to a narrow ($\sim$1 keV broad, red
curve) component with a negative mass of $- 0.05~M_\odot$, which can
be interpreted as a marginal evidence for a dip in the velocity
substructure, given that this improvement of the $\Delta \chi^2$
  came at the cost of adding three more parameters\footnote{We note,
    that the width and especially energy of the reference line are
    very nonlinear parameter that could lead to large changes in the
    $\chi^2$.} to the model. One can estimate the constraints on the
  line flux (mass of a transparent clump) that such
  analysis can provide, by fixing the centroid energy and the width of the
  reference 847 keV line and calculating the expected statistical
  uncertainty. Since the normalization of the \tem model is the only
free parameter in this particular experiment, the estimation of the
uncertainty is straightforward (see Fig.\ref{fig:sigma_m}). Three
curves shown in Fig.\ref{fig:sigma_m} show 1$\sigma$ uncertainty on
the initial $^{56}$Ni mass for the \early set (dashed-blue: SPI data
in the 70-1350 keV band; long-dashed-green: 400-1350 keV) and \late
set (solid-red: 400-1350 keV), respectively. Conservative upper limit
based on the assumption of pure statistical errors would be 3 times
these values. Letting the broadening and the redshift to be free
parameters (look-elsewhere effect) would increase this limit even
further.
 
These experiments show that the \late data are consistent with a
presence of a velocity substructure (parameterized via our \tem
model) on top of the 1D \W7 model at the level $\sim 0.05~M_\odot$,
provided that the lines are slightly broadenened. 

\begin{figure*}
\begin{center}
\includegraphics[trim= 0cm 5cm 0cm 2cm,
  width=1\textwidth,clip=t,angle=0.,scale=0.49]{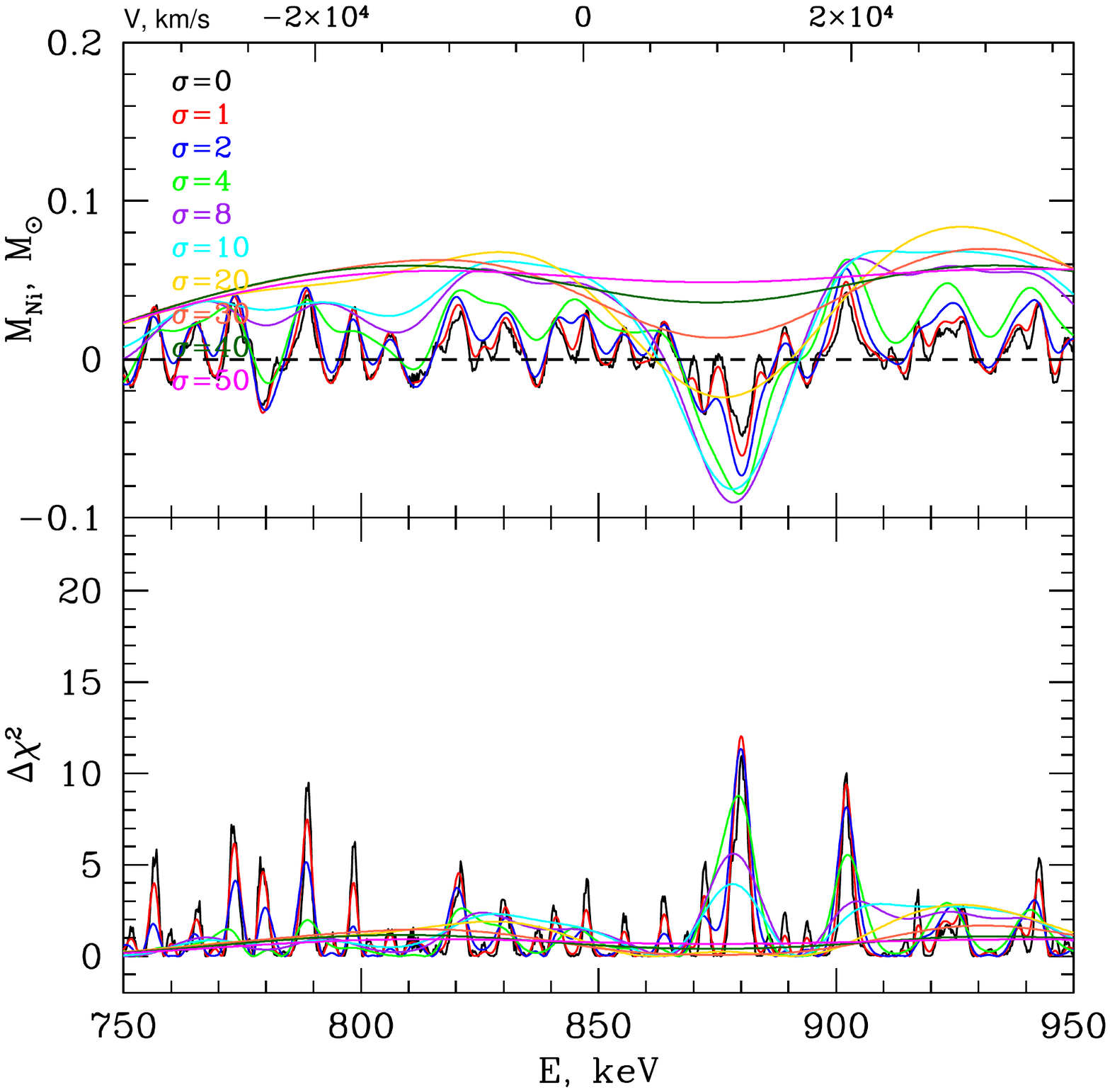}
\includegraphics[trim= 0cm 5cm 0cm 2cm,
  width=1\textwidth,clip=t,angle=0.,scale=0.49]{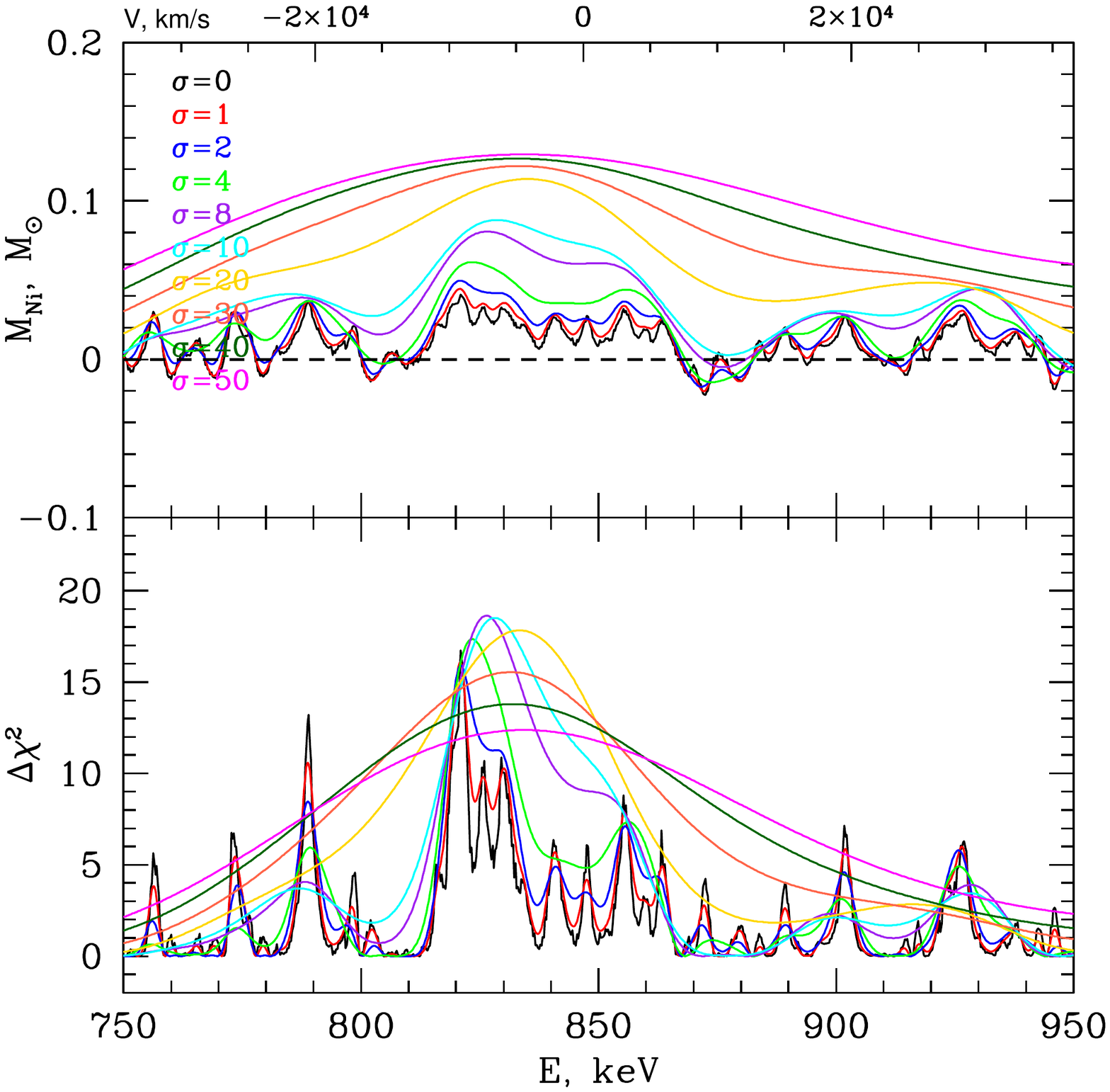}
\end{center}
\caption{Same as in Fig.\ref{fig:tem_late} but for the \early SPI
  spectrum. {\bf LEFT:} \tem+\W7 model and the SPI data in
  the 400-1350 keV band. The normalization of the \W7 model is fixed to
  1. {\bf RIGHT:} \tem+\W7 model and the SPI data in
  the 70-1350 keV band. The normalization of the \W7 model is fixed to
  1. The low energy part of the SPI spectrum is included to make sure
  that the $^{56}$Ni line at 158 keV is within the energy range
  probed. There is a marginal evidence of a redshifted (by $\sim$8000 km/s)
  component with the width of $\sim 8$ keV (Gaussian sigma),
  corresponding to $M_{Ni}\sim0.08~M_\odot$. See text
  for the discussion. 
\label{fig:tem_early}}
\end{figure*}

We now do a similar experiment with the \early data, using a \tem+\W7
model for SPI data in the 400-1350 keV band (Fig.\ref{fig:tem_early},
left panel) and in the 70-1350 keV band (Fig.\ref{fig:tem_early},
right panel), respectively. 

The left panel does not show any significance evidence for a clump on
top of the \W7 model.  The structure in the right panel is more
complicated. The data used in this panel now include the $^{56}$Ni
line at 158 keV. We note, that if the 158 keV line is able to escape,
then it is certainly true for higher energy lines of Ni and
Co. Therefore the analysis should be done for the whole band to
achieve the most significant results.  First of all, our analysis does
not show compelling evidence for a narrow and unshifted component
reported in \citet{2014Sci...345.1162D} -- there is a weak
($\Delta\chi^2\sim6$, i.e. $\sim 2.4~\sigma$ detection if we ignore the freedom in the redshift and broadening) peak at 847.5 keV,
corresponding to a narrow line (black curve) with a mass of $\sim
0.027~M_\odot$ of $^{56}$Ni. There are several separate peaks of
similar magnitudes, covering the energy range of interest.  However
there is a more significant (albeit also marginal) evidence for a
redshifted and broad component with $M_{Ni}\sim0.08~M_\odot$,
$E\sim826.5~$keV and $\sigma\sim8~$keV \citep[see][for
  discussion]{isern}.  The gain in $\chi^2$ is $\sim$18 and for a
fixed energy and broadening (putting under the rug possible systematic
errors in the background modeling and uncertainties in the calibration
of the off-diagonal response) this would be a 4.2$~\sigma$ detection. However
the freedom in the energy, width (look elsewhere effect) and the
normalization deteriorates the significance. Should all these free
parameters be linear (as is normalization), one would expect the
change in the $\chi^2$ of $\sim 3$ due to pure statistical
fluctuations. However, the energy and the width are nonlinear and the
gain in $\chi^2$ might be significantly larger. In
Fig.~\ref{fig:tem_late} and \ref{fig:tem_early} we see multiple peaks
with the change/gain in $\chi^2$ up to $\sim$10. Assuming that the
latter value can be used as a crude estimate of a possible gain in
$\chi^2$ due to non-linearity of the \tem model, the significance of
the detection of the excess drops below 3$\sigma$.

Taking the best-fitting parameters at the face value, we can go back
to the \late data and compare the spectra (in the 400-1350 keV band)
with the \tem+\W7 model, freezing \tem model parameters at the
best-fitting values obtained for the \early data. This gives the
$\chi^2=1883.05$, i.e. worse than the \W7 model alone
($\chi^2=1879.3$). If we let the normalizations of both \tem and \W7
models free (but freezing energy and broadening of the \tem model),
then we improve slightly the $\chi^2$ to 1878.9, but the best-fitting
mass becomes slightly negative, although consistent with zero $-3~10^{-3}\pm5~10^{-2}~M_\odot$, while the
best-fitting normalization of \W7 model becomes 0.92
(c.f. Tab.~\ref{tab:mfit_late} where SPI data are used together with
the ISGRI data).

We concluded that there is a tension between ``low'' energy SPI data
in \early observations and the rest of the \INTEGRAL data
(Tab.~\ref{tab:mfit_early} and Tab.~\ref{tab:lc_chi2}). However, this tension is not prohibitively large and could be attributed to statistical fluctuations in the data, if a conservative approach is adopted. A possible
evidence of the redshifted and broadened 158 keV line in the \early
data and possible implications are further discussed in \citet{isern}.

\subsection{3PAR model}
\label{sec:3par}
Apart from the models discussed above, we also used \PAR3 model,
introduced in \citet{2014Natur.512..406C}. This is a spherically
symmetric model of homologously expanding ejecta with exponential
density profile $\displaystyle \rho \propto e^{v/V_e}$. The model is
chaaracterized by three parameters: initial mass of the $^{56}$Ni
$M_{Ni}$, total mass of the ejecta $M_{ejecta}$, and characteristic
expansion velocity $V_e$ in the exponential density distribution. In
this model a mass-weighted root-mean-squared velocity of the ejecta is
$\displaystyle \sqrt{12}V_e$.

The main shortcoming of this model is the assumption that all
elements, including radioactive Ni and Co, are uniformly mixed through
the entire ejecta. This is an ad-hoc assumption, made in order to stay
with only three-parameteres model, but it is not justified. It has the
major impact for the early gamma-ray light curve, producing gamma-ray
emission even at the very early phase (see Fig.~\ref{fig:background},
\ref{fig:lc_isgri}, \ref{fig:lc_spi}). At later times (day 50 or
later), the role of mixing is less significant. We therefore applied
this model to the \late ISGRI and SPI spectra to get estimates of
$M_{Ni}$, $M_{ejecta}$ and $V_e$, which are not limited to values characteristic to the set of plausible models given in Table~\ref{tab:models}. The main purpose of using this model is to understand the level of constraints provided by the \INTEGRAL data on the main characteristics of the supernova. Simplicity of the model allows us to calculate this model on a large grid of possible values of  $M_{Ni}$, $M_{ejecta}$ and $V_e$.

A Monte Carlo radiative transfer code is used to calculate the
emergent spectrum, which includes full treatment of Compton scattering
(coherent and incoherent) and photoabsorption. Pair production by
$\gamma$-ray photons is neglected. The positrons produced by $\beta^+$ decay
annihilate in place via positronium formation. Both two-photon
annihilation into the 511 keV line and the ortho-positronium continuum
are included.

\begin{figure*}
\begin{center}
\includegraphics[trim= 0cm 12cm 0cm 1cm,
  width=1\textwidth,clip=t,angle=0.,scale=0.99]{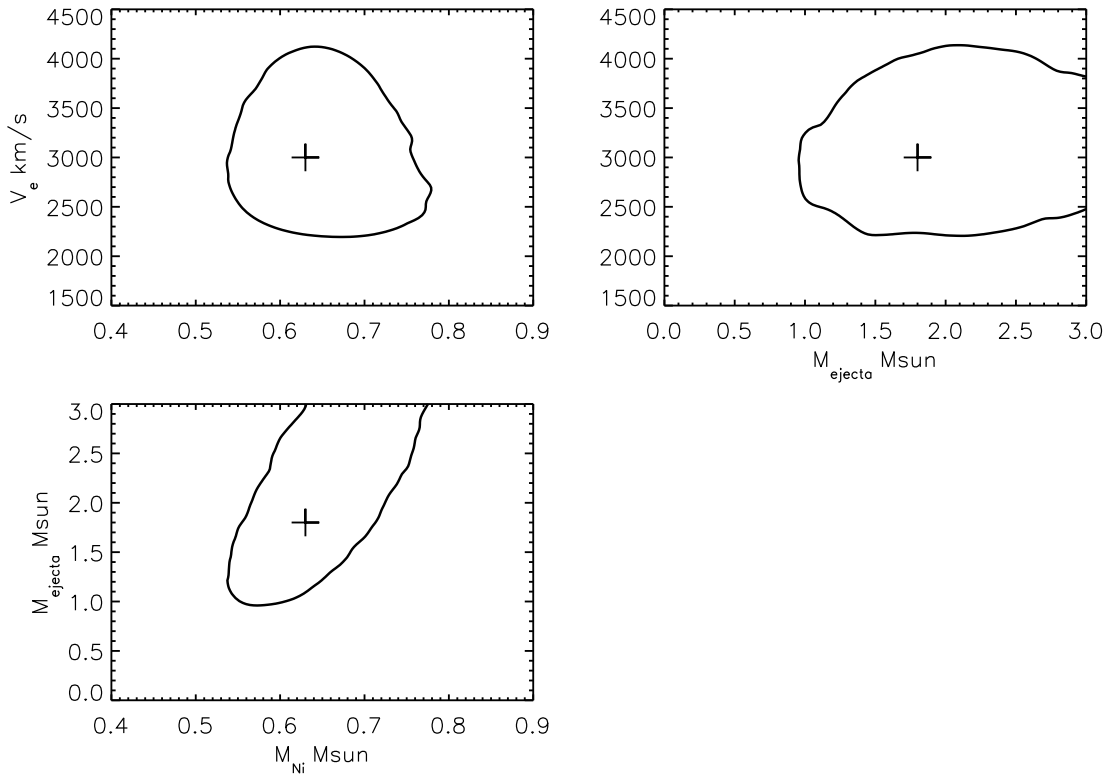}
\end{center}
\caption{Confidence contours for \PAR3 model, corresponding to $\Delta\chi^2=1$ with respect to the best-fitting value. The cross show the best-fitting parameters of the \PAR3 model: $M_{Ni}\sim 0.63~M_\odot$, $v_e\sim 3000~{\rm km~s^{-1}}$, $M_{ejecta}\sim 1.8~M_\odot$. The \late ISGRI and SPI spectra are used for this analysis.
Confidence intervals plotted in this figure correspond to $1~\sigma$ for a single parameter of interest.
The largest uncertainty is in the mass of the ejecta, while the Ni mass is the best determined quantity.
\label{fig:3par}}
\end{figure*}

The results are shown in Fig.~\ref{fig:3par}. The best-fitting values
$M_{Ni}=0.63~M_\odot$, $M_{ejecta}=1.8~M_\odot$, $V_{e}=3~10^{3}~{\rm
  km~s^{-1}}$ are marked with a cross. The 1$\sigma$ confidence
contours (corresponding to $\Delta \chi^2=1$, i.e. for single
parameter of interest) are shown with the thick solid line. Clearly,
the Ni mass $M_{Ni}$ and the characteristic expansion velocity $V_e$
are better constrained than the ejecta mass. This is not surprizing,
given that the data averaged over the period 50-162 days after
explosion are used, when the ejecta are relatively transparent for
gamma-rays. As a results the flux in the lines depends primarily on
the Ni mass, line broadening is set by the expansion velocity, while
ejecta mass influence mostly the amplitude of the scattered component,
which declines with time relative to the ortho-positronium continuum
when the optical depth declines. If we fix the poorly constrained
ejecta mass to $M_{ejecta}=1.4~M_\odot$, then the derived Ni mass is
constrained to the range 0.54-0.67$~M_\odot$.

For the set of models listed in the Table~\ref{tab:models} we can
estimate the effective $V_e$ using the relation $\displaystyle
V_e=\sqrt{\frac{E_K}{6M_{ejecta}}}$, valid for pure exponential
model. The values $\displaystyle V_e$ vary between $\sim2580~{\rm
  km~s^{-1}}$ for \DDTe  to $\sim2960~{\rm km~s^{-1}}$ for \DETO models
and is equal to $2740~{\rm km~s^{-1}}$ and $2820~{\rm km~s^{-1}}$ for
\W7 and \texttt{DDT1p1} respectively. Not surprisingly all
``successful'' models (e.g. \W7 and \texttt{DDT1p1}) have their
characteristic parameters well inside contours plotted in
Fig.~\ref{fig:3par}, while \DETO and \HED6 are far outside the
contours, primarily because of Ni mass.

\subsection{Summary of model fitting}
\label{sec:mf_sum}
The comparison of the \INTEGRAL data with the subset of models (see the
sections above) allows one to crudely rank the models according to
their success in different tests. For each test (data set) we can
choose the ``best'' model, which provides the largest improvement
$\Delta\chi^2$ compared to Null model (or having the smallest $\chi^2$
for the lightcurves). We can then adopt an ad hoc definition that
other models that have $\chi^2$ different from the best model by 4
(i.e. $\sim 2~\sigma$ confidence) are classified as ``good''. Similar approach can be applied to the lightcurves in each band (Tab. \ref{tab:lc_chi2}), by adding 4 to the minimal value of the $\chi^2$ among models. Applying
this test to Tables \ref{tab:mfit_late} - \ref{tab:lc_chi2} we
conclude that \W7 and \texttt{DDT1p1} pass all these tests, closely followed by
\DD4, \texttt{ddt1p4}, and then by \texttt{DDT1p4halo} and \BALL3. \DETO and \HED6 fail most of the tests. Of course, given
the uncertainties in the distance, background modeling and calibration
issues, we can not reject models other than \DETO and \HED6. E.g., if we let the
normalization to be a free parameter (equivalent of a statement that
the distance is highly uncertain) then most of the models become
barely distinguishable.  We rather state, that a whole
class of near-Chandrasekhar models provides a reasonable description
of the data, with the \W7 and \texttt{DDT1p1} being the most successful, closely followed
by a broader group of delayed-detonation models.

\section{Consistency with optical data}
\label{sec:opt}
We now make several basic consistency checks of gamma-ray and optical
data, using optical observations taken quasi-simultaneously with
\INTEGRAL observations.

\subsection{Optical and gamma-ray luminosities}
We use $BVRIJHK$ photometry reported by \citet{2014MNRAS.443.2887F} to
estimate the bolometric (UVOIR) luminosity of SN~2014J on days 73 and
96 after the explosion. Since the data do not contain the $U$-band
photometry, we include the $U$ magnitude recovered on the bases of the
$U-B$ color of the dereddened normal SN~Ia, SN~2003hv
\citep{2009A&A...505..265L}.  The SN~2014J fluxes were corrected for
the extinction using slightly different extinction laws reported by
\citet{2014ApJ...788L..21A} and \citet{2014MNRAS.443.2887F}.  The
average of both fluxes for each epoch were used then to produce the
integrated flux.  To this end we approximated the spectral energy
distribution by the combination of two functions each of which is a
smooth broken power law. The SED integration in the range of
$0.1<\lambda<10$ $\mu$m with the distance of 3.5 Mpc, results in the
luminosity estimates of $(11\pm1)\times10^{41}$ erg s$^{-1}$ on day
73, and $(6.5\pm0.6)\times10^{41}$ erg s$^{-1}$ on day 96.  These
values agree well with the estimated amount of deposited energy in
the best-fitting \PAR3 model: $\sim 1.0\times 10^{42}~{\rm erg~s^{-1}}$
and $\sim 5.3\times~10^{41}~{\rm erg~s^{-1}}$ for day 73 and 96
respectively. According to this model the fraction of thermalized
energy is $\sim$34\% and $\sim$20\% for these dates respectively.

\subsection{Asymmetry in late optical spectra?}

The issue of asymmetry of SN~2014J ejecta is of vital importance
because the strong deviation of the $^{56}$Ni distribution from the
spherical symmetry would affect the interpretation of the gamma-ray
data.  Generally, the asymmetry of the $^{56}$Ni distribution is
expected in the binary WD merger scenario
\citep{2012ApJ...747L..10P}. Moreover, a single degenerate scenario
also does not rule out the ejecta asymmetry caused by the noncentral
early deflagration \citep{2014ApJ...782...11M}.  In fact, signatures
of asymmetry have been already detected in several SNe~Ia at the
nebular stage ($t > 100$ d). The asymmetry is manifested in the
emission line shift or/and the double peak emission line profiles
\citep{2006ApJ...652L.101M,2010Natur.466...82M,2014arXiv1401.3347D}.

\begin{figure}
\begin{center}
\includegraphics[trim= 0cm 2cm 0cm 0cm,
  width=1\textwidth,clip=t,angle=0.,scale=0.5]{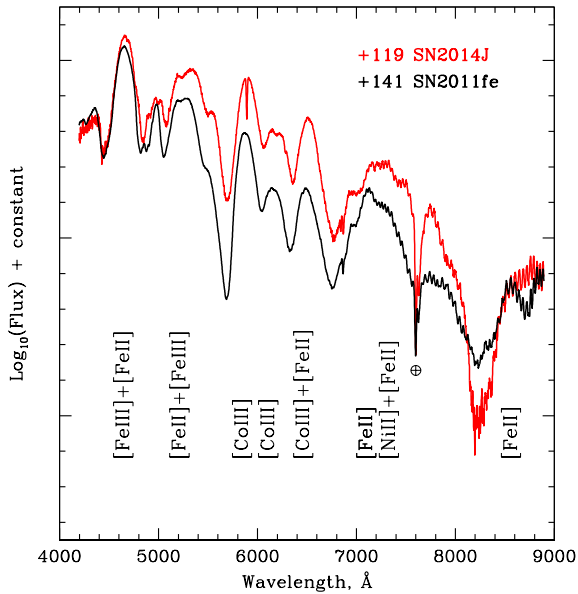}
\end{center}
\caption{The spectra of SN~2014J (day 119 after the maximum) and 
SN2011fe (day 141 after the maximum) obtained with RTT-150 
telescope \cite{bikmaev15}. Overall the spectra are very similar in 
terms of the flux level, line shape
and line ratios. The exception is the prominent blueshift of [Fe\,III], 
[Fe\,II], and [Co\,III] emissions of SN~2011fe relative to SN~2014J. 
The strong interstellar Na\,I absorption in the SN~2014J spectrum 
arises in the M~82 galaxy.
\label{fig:rtt-sn2011fe}}
\end{figure}

To probe a possible asymmetry of SN~2014J ejecta we rely on the
nebular optical spectrum taken on day 119 after the $B$ maximum, i.e.,
136 d after the explosion \citep{bikmaev15} at the 1.5-m
Russian-Turkish telescope (RTT-150) of the TUBITAK National
Observatory (Antalya, Turkey).  The SN~2014J spectrum corrected for
the interstellar reddening in M82 of $E(B-V) = 1$
\citep[c.f.][]{2014MNRAS.443.2887F} is shown in
Fig.~\ref{fig:rtt-sn2011fe} together with that of SN~2011fe obtained
at the same instrument on day 141 after the maximum. The spectra of
both supernovae look similar except for the blueshift of SN~2011fe
emissions by $\sim 10^3$ km s$^{-1}$ relative to SN~2014J.

\begin{figure}
\begin{center}
\includegraphics[trim= 5cm 8cm 0cm 4cm,
  width=1\textwidth,clip=t,angle=0.,scale=0.6]{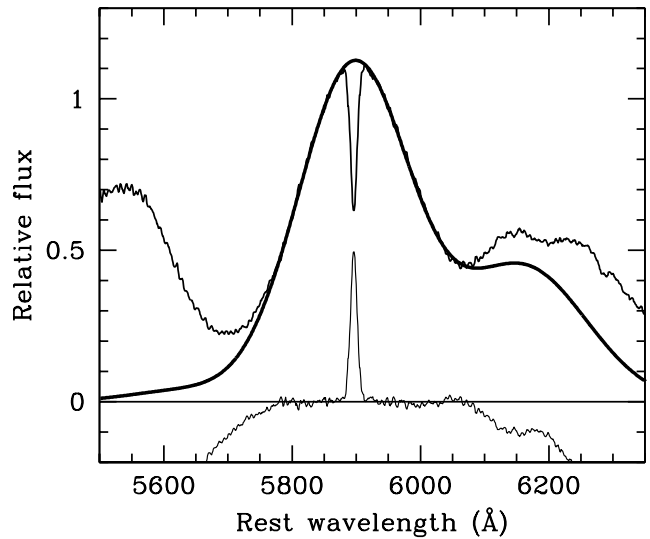}
\end{center}
\caption{[Co\,III]  5900 \AA\ emission in the SN2014J spectrum on day 119
({\bf thin} line) along with the model ({\bf thick} line) which 
includes five component of the a$^4$F - a$^2$G multiplet.
The narrow absorption feature at the top of the profile is due 
to Na\,I  interstellar absorption in M 82. At the bottom shown is 
the residual "model minus observation", which demonstrates a good 
fit in the range of 5770-6060 \AA.
\label{fig:fco3}}
\end{figure}

We focus on the [Co\,III] 5890 \AA\ emission that is not hampered 
markedly by the blending with other lines. It should be emphasised 
that on day 136 d after the explosion this line is dominated by $^{56}$Co; 
the contribution of $^{57}$Co and stable Co isotopes is negligible. 
The Thomson optical depth at this epoch is small ($\sim0.2$) and 
does not affect the line profile. 
The [Co\,III] emission is the superposition of 
five lines of the a$^4$F - a$^2$G multiplet. Each line we describe by 
the Gaussian with the amplitude proportional to the 
collisional excitation rate times the radiative branching ratio. 
We adopt the heliocentric recession 
velocity of $+104\pm15$ km s$^{-1}$ that takes into account the 
recession velocity of +203 km s$^{-1}$ for M 82 (NASA Extragalactic 
Database NED) and the rotational velocity of M 82 at the SN~2014J position.
The best fit (Fig. \ref{fig:fco3}) is found for the full width at half 
maximum for each line FWHM = 10450 km s$^{-1}$ and the line shift of 
$v_s=+130\pm17$ km s$^{-1}$. 
With the exception of this small shift, each [Co\,III] line 
is fairly symmetric at least in the radial velocity range of 
$|v_r| < 6100$ km s$^{-1}$.
The small line shift may be related to either 
intrinsically small asymmetry of $^{56}$Ni distribution, or the special
viewing angle, if the ejecta is actually non-spherical.
To summarize, the SN~2014J optical spectrum does not show signatures
of strong asymmetry.

%and, therefore, do not provide any direct
%support for merger scenario.

%===============================================================

\section{Discussion and conclusions}
\label{sec:conclusions}
We have analyzed a complete set of \INTEGRAL observations of
SN2014J. We confirm our previous results \citep{2014Natur.512..406C}
that the data are broadly consistent with the predictions of a
nearly-Chandrasekhar WD explosion, with (1D) deflagration or delayed
detonation models providing equally good description (see Tables
\ref{tab:mfit_eandl} - \ref{tab:lc_chi2}).  While pure deflagration
models are disfavored because of the expected large scale mixing and
incomplete burning in 3D simulations, in the 1D case they yield the
same gamma-ray flux as the delayed detonation models. Pure detonation
(or strongly sub-Chandrasekhar) models strongly overproduce
(underproduce) observed gamma-ray flux and can be excluded. Allowing a
freedom in the normalization of the model (equivalent to allowing the
initial mass of $^{56}$Ni to be a free parameter, while keeping other
parameters unchanged) makes all models essentially indistinguishable
at the level of statistics, accumulated by \INTEGRAL.

We have searched for possible velocity substructure on top of the
predictions from 1D models, by adding a set of broadened Gaussian
lines to the best-performing \W7 model. The energies and fluxes of the
lines are tied to the predictions of the Ni and Co decay chains,
appropriate for the optically thin clump of Ni. This analysis did not
reveal strong evidence for a prominent velocity substructure in the
gamma-ray data during the late phase of the SN evolution (after day
50). Given the statistics accumulated by {\it INTEGRAL}, a clump with
the $^{56}$Ni mass $\sim 0.05~M_\odot$ producing slightly broadened
lines (Fig.\ref{fig:sigma_m}) could be consistent with the \late
gamma-ray data. Similar analysis of the \early data has a best-fitting
solution with a redshifted and broadened component with
$M_{Ni}\sim0.08~M_\odot$, $E\sim826.5~$keV and
$\sigma\sim8~$keV. However, the statistical significance of this extra
component is marginal and the \late observations do not provide
further evidence for the presence of such component \citep[see
  also][for independent analysis of early observations of
  SN2014J]{2014Sci...345.1162D,isern}.

From the optical light curves and spectra SN2014J appears to be a
``normal'' SNIa with layered structure and no evidence for large-scale
mixing \citep[e.g.,][]{2015ApJ...798...39M,2014MNRAS.445.4427A},
consistent with the delayed-detonation models.  The detection of
stable Ni \citep{2014ApJ...792..120F,2015ApJ...798...93T} in IR
suggests high density of the burning material, characteristic for
near-Chandrasekhar WD.

Optical spectrum taken at the nebular stage (day $\sim 136$ after the
explosion) also do not show strong asymmetry in the Co and Fe
lines. Unless the viewing angle is special, the distribution of these
elements in the ejecta is symmetric.  These data do not provide any
direct support for collision/merger scenario. The late SN2014J
spectrum is very similar to that of SN2011fe, albeit with the
pronounced blueshift of emission lines of the latter.

Apart from the above mentioned feature in the \early observation,
which we consider as marginal, the rest of the \INTEGRAL and optical
data appear consistent with the predictions of ``canonical'' 1D
explosion models of a nearly-Chandrasekhar carbon-oxygen white dwarf.

\vspace*{1cm}

\acknowledgments

This work was based on observations with INTEGRAL, an ESA project with 
instruments and a science data centre funded by ESA member states 
(especially the principal investigator countries: Denmark, France, 
Germany, Italy, Switzerland and Spain) and with the participation of 
Russia and the United States. We are grateful to ISOC for their scheduling efforts, and the INTEGRAL Users Group for their support in the observations.
E.C., R.S., S.G.  are partly supported by 
grant No. 14-22-00271 from the Russian Scientific Foundation. J.I. is 
supported by MINECO-FEDER and Generalitat de Catalunya grants. I.B. is partly supported by Russian Government Program of Competitive Growth of KFU. E.B. is supported by Spanish MINECO grant AYA2013-40545. 
The SPI project has been completed under the responsibility and 
leadership of CNES, France. ISGRI has been realized by CEA with the 
support of CNES. We thank Adam Burrows, Peter H\"oflich, Rishi Khatri, Ken Nomoto, 
Victor Utrobin, Alexey Vikhlinin and Stan Woosley for helpful discussions.

\appendix
\section{Comparison of simple models via $\Delta\chi^2$ criterion}
In this appendix we clarify our approach of comparing different models
using ungrouped spectral and/or timing data.

\subsection{Grouping the data and small number of counts per bin}
Consider $N$ data points $D_i$, $i=1,N$ and $N\gg 1$ (e.g., a spectrum measured in $N$ energy bins) that correspond to a model
$M_{t}$, $D_i=M_{t,i}+n_i$, where $n_i$ is the noise with Gaussian
distribution with zero mean and known variance $\sigma_i^2$. The noise is
uncorrelated, i.e. $<n_in_j>=0$, if $i\ne j$. For simplicity we drop
below the index $i$ in the expressions containing summation over $i$.

Suppose that we want to compare two competing models $M_1$ and $M_2$
with no free parameters (this corresponds to Neyman-Pearson lemma of
two simple hypotheses). For our purpose it is useful to write
explicitly the probability distribution of $\Delta \chi^2$ between
competing models.

Let us define $\delta_{1}=M_1-M_t$ and $\delta_{2}=M_2-M_t$ and
calculate $\chi^2$ for both models. For the model $M_1$
\begin{eqnarray}
  \chi_1^2=\sum{\left (\frac{M_1-D}{\sigma} \right )^2}=\sum{\left (\frac{\delta_1+n}{\sigma} \right )^2}=\sum{\left (\frac{\delta_1}{\sigma} \right )^2}+2\sum{\frac{\delta_1n}{\sigma^2}}+\sum{\left (\frac{n}{\sigma} \right )^2},
\end{eqnarray}
and a similar expression for $M_2$. Thus $\Delta \chi^2=\chi^2_1-\chi^2_2$ is
\begin{eqnarray}
  \Delta \chi^2= \sum{\left (\frac{\delta_1}{\sigma} \right )^2}-\sum{\left (\frac{\delta_2}{\sigma} \right )^2} + 2\sum{\frac{\delta_1n}{\sigma^2}}-2\sum{\frac{\delta_2n}{\sigma^2}}=\\
  \sum{\left (\frac{\delta_1}{\sigma} \right )^2}-\sum{\left (\frac{\delta_2}{\sigma} \right )^2} + 2\sum{\frac{\delta_{1,2}n}{\sigma^2}},
\end{eqnarray}
where $\delta_{1,2}=M_1-M_2$.  The last term in the above expression
obviously has a Gaussian distribution with zero mean and variance
$\displaystyle 4
\sum{\frac{\delta_{1,2}^2<n^2>}{\sigma^4}}=4\sum{\frac{\delta_{1,2}^2}{\sigma^2}}$. Thus
\begin{eqnarray}
\Delta \chi^2=\sum{\left (\frac{\delta_1}{\sigma} \right )^2}-\sum{\left (\frac{\delta_2}{\sigma} \right )^2} + X\sqrt{4\sum{\frac{\delta_{1,2}^2}{\sigma^2}}},
\end{eqnarray}
where $X$ has a Normal distribution. It is clear that the above expression 
{\it does not depend on data grouping} (see Churazov et al., 1996) as long as the grouping does not severely affect the shapes of $\delta_1$, $\delta_2$ or $\delta_{1,2}$. Furthermore, $\Delta
\chi^2$ can have a Gaussian distribution by the central limit theorem,
even when the noise in the original data is not Gaussian. For
instance, if $n$ corresponds to the Poisson noise in the data with
small number of counts per bin, the  $\Delta
\chi^2$ will have Gaussian distribution provided that the {\it total
  number of counts} contributing to
$\sum{\frac{\delta_{1,2}n}{\sigma^2}}$ is large (Churazov et al.,
1996). We therefore can use the original data with no grouping to calculate $\Delta \chi^2$. 

\subsection{Dividing the models into a ``more plausible'' and ``less plausible'' groups}
For a given observed $\Delta \chi^2_{obs}$ the ratio of likelihoods
for two models is $\displaystyle \Lambda=\frac{L(M_1)}{L(M_2)}=e^{-\Delta
  \chi^2_{obs}/2}$, which corresponds to the Akaike information
criterion (AIC) or Bayesian information criterion (BIC)
criteria. Alternatively we can also employ Neyman-Pearson lemma to
differentiate between two models. If $M_1$ is the true model, then
$M_1=M_t$, $\delta_1=0$, $\delta_2=-\delta_{1,2}$ and the distribution
of $\Delta \chi^2=\chi^2_1-\chi^2_2$ is
\begin{eqnarray}
\Delta \chi^2=-\sum{\left (\frac{\delta_{1,2}}{\sigma} \right )^2}+ X\sqrt{4\sum{\frac{\delta_{1,2}^2}{\sigma^2}}}.
\end{eqnarray}
Then the probability $\alpha$ of getting $\Delta \chi^2 > y$ is
\begin{eqnarray}
  \alpha=Q(x),
  \label{eq:alpha}
\end{eqnarray}
where $Q(x)=1/2 ~{\rm erfc}(x/\sqrt{2})$, 
$\displaystyle x=\frac{A+y}{2\sqrt{A}}$ and $A=\sum{\left
  (\frac{\delta_{1,2}}{\sigma} \right )^2}$. Although the value of $A$ is
known, we can take a conservative approach and write that
$\frac{A+y}{2\sqrt{A}}\ge \sqrt{y}$ for any $A \in [0,\infty]$. This
value is achieved at $A=y$. Thus one can conclude that the
conservative estimate of the probability of getting better $\chi^2$
for the wrong model (i.e., $\Delta \chi^2_{obs} > y$) corresponds to more
than $\sqrt{y}$ standard deviations. In the paper we use the value of
$\Delta \chi^2_{obs} < 4$ to separate the models into a ``more plausible''
and ``less plausible'' groups. As we emphasized above, one can also
interpret this value in the frame of BIC or AIC approaches.

\subsection{Pairwise model comparison and the goodness of fit criterion}
When comparing two models in terms of the $\chi^2$ we are effectively
projecting the N-dimensional data $D$ on a line connecting the
models $M_1$ and $M_2$. We can formally introduce a linear parameter
$p$, make a composite model $M_1+p(M_2-M_1)$
and calculate best-fitting value $p_{bf}$ that minimizes the $\chi^2$
\begin{eqnarray}
p_{bf}=\frac{\sum -(D-M_1)\delta_{1,2}/\sigma^2}{\sum \delta_{1,2}^2/\sigma^2}.
\end{eqnarray}
It is obvious that this quantity has a Gaussian distribution with the standard deviation
\begin{eqnarray}
\sigma_p=\left( \sum \delta_{1,2}^2/\sigma^2\right )^{-1/2}.
\end{eqnarray}
We can now use the deviation of $p_{bf}$ from 0 or 1 as a goodness of
fit criterion for models $M_1$ and $M_2$, respectively. A significant
deviation (i.e. $|p_{bf}/\sigma_p|\gg 1$ or $|(p_{bf}-1)/\sigma_p|\gg
1$) implies that one of the models (or both) is unlikely. While the
goodness of fit can be calculated for each model alone, the power of
the criterion depends on the data grouping (imagine, for example, a
very weak signal distributed over a large number of bins). The
goodness of fit calculation described above verifies only the
projection of the data on the line set by models, but it
``optimally'' compares the difference between plausible models with
the signal present in the data.

For each model used in \S\ref{sec:models} we have calculated the
maximum deviation $|p_{bf}/\sigma_p|$ with respect to all other models
(except for the \BALL3 model that was designed to fit the \INTEGRAL
data). Table \ref{tab:m12} provides corresponding values, when ISGRI
and SPI data for the \early and \late periods are considered
jointly. Based on this analysis we conclude that \texttt{No source},
\DETO, \HED6 and (marginally) \DDTe models are disfavored by the data,
while other models are compatible with the data.

\begin{deluxetable*}{lrr}
\tabletypesize{\footnotesize}
\tablecaption{Pairwise comparison of models from Table~\ref{tab:models}. For each model the maximum deviation $|p_{bf}/\sigma_p|$ with respect to all other models is given.  
}
\tablewidth{12cm}
\tablehead{
\colhead{} &
\colhead{ISGRI \& SPI(400-1350 keV)} &
\colhead{ISGRI \& SPI(70-1350 keV)}\\
\colhead{Model} &
\colhead{Maximum deviation } &
\colhead{Maximum deviation} 
}
\startdata
\texttt{No source}  &      9.0  &     11.1 \\ 
\texttt{DDT1p1}     &      0.7  &      1.8 \\ 
\texttt{DDT1p4halo} &      1.1  &      2.1 \\ 
\DDTe               &      2.2  &      3.6 \\ 
\DETO               &     12.5  &     12.6 \\ 
\HED6               &      4.3  &      5.4 \\ 
\W7                 &      0.8  &      1.3 \\ 
\texttt{ddt1p4}     &      2.3  &      1.3 \\ 
\DD4                &      1.6  &      1.0  
\enddata
\tablecomments{The ISGRI and SPI data for the \early and \late period are used jointly. Large deviations imply that the model is disfavored by the data. Using a threshold of 2 standard deviations one can conclude that \texttt{No source}, \DETO, \HED6 and (marginally) \DDTe models are disfavored by the data, while other models are compatible with the data. Since \BALL3 model that was designed to fit the \INTEGRAL data in the \early period, this model has been excluded from this test.}
\label{tab:m12}
\end{deluxetable*}

\bibliographystyle{apj}

\begin{thebibliography}{99}

\bibitem[Achtermann \& Lacy(1982)]{1995ApJ...439..163A} Achtermann, J. M. \& 
 Lacy, J. H.\ 1995, ApJ, 439, 163

\bibitem[Arnett(1982)]{1982ApJ...253..785A} Arnett, W.~D.\ 1982, \apj, 253, 
785 
\bibitem[Amanullah et al.(2014)]{2014ApJ...788L..21A} Amanullah, R., 
Goobar, A., Johansson, J., et al.\ 2014, \apjl, 788, L21 

\bibitem[Ambwani 
\& Sutherland(1988)]{1988ApJ...325..820A} Ambwani, K., \& Sutherland, P.\ 1988, \apj, 325, 820 

%\bibitem[Ashall et al.(2014)]{2014arXiv1409.7066A} Ashall, C., Mazzali, P., 
%Bersier, D., et al.\ 2014, arXiv:1409.7066 
\bibitem[Ashall et al.(2014)]{2014MNRAS.445.4427A} Ashall, C., Mazzali, P., 
Bersier, D., et al.\ 2014, \mnras, 445, 4427 

\bibitem[Axelrod(1980)]{1980PhDT.........1A} Axelrod, T. S. 1980, Ph.D. Thesis California Univ., Santa Cruz.

\bibitem[Bachetti et al.(2014)]{2014Natur.514..202B} Bachetti, M., 
Harrison, F.~A., Walton, D.~J., et al.\ 2014, \nat, 514, 202 


\bibitem[Badenes et al.(2003)]{2003ApJ...593..358B} Badenes, C., Bravo, E., 
Borkowski, K.~J., \& Dom{\'{\i}}nguez, I.\ 2003, \apj, 593, 358 


\bibitem[Bikmaev et 
al.(2015)]{bikmaev15} Bikmaev, I.,  et al.\ 2015, (Astronomy Letters, submitted)

\bibitem[Blondin et al.(2013)]{2013MNRAS.429.2127B}
Blondin, S., Dessart, L., Hillier, D.~J., Khokhlov, A.~M. 2013, MNRAS, 
   429, 2127

%\bibitem[Brown et al.(2014)]{2014arXiv1408.2381B} Brown, P.~J., Smitka, 
%M.~T., Wang, L., et al.\ 2014, arXiv:1408.2381 
\bibitem[Brown et al.(2015)]{2015ApJ...805...74B} Brown, P.~J., Smitka, 
M.~T., Wang, L., et al.\ 2015, \apj, 805, 74 

\bibitem[Chugai(2000)]{2000AstL...26..797C} Chugai, N.~N. 2000, AstL, 
   26, 797

\bibitem[Churazov et al.(1996)]{1996ApJ...471..673C} Churazov, E., 
Gilfanov, M., Forman, W., \& Jones, C.\ 1996, \apj, 471, 673 


\bibitem[Churazov et al.(2011)]{2011MNRAS.411.1727C} Churazov, E., Sazonov, 
S., Tsygankov, S., Sunyaev, R., 
\& Varshalovich, D.\ 2011, \mnras, 411, 1727 


\bibitem[Churazov et al.(2014a)]{2014ATel.5992....1C} Churazov, E., Sunyaev, 
R., Grebenev, S., et al.\ 2014a, The Astronomer's Telegram, 5992, 1 


\bibitem[Churazov et al.(2005)]{2005MNRAS.357.1377C} Churazov, E., Sunyaev, 
R., Sazonov, S., Revnivtsev, M., 
\& Varshalovich, D.\ 2005, \mnras, 357, 1377 

\bibitem[Churazov et al.(2014b)]{2014Natur.512..406C} Churazov, E., Sunyaev, 
R., Isern, J., et al.\ 2014b, \nat, 512, 406 

\bibitem[Clayton et al.(1969)]{1969ApJ...155...75C} Clayton, D.~D., 
Colgate, S.~A., \& Fishman, G.~J.\ 1969, \apj, 155, 75 

\bibitem[Diehl et al.(2014)]{2014Sci...345.1162D} Diehl, R., Siegert, T., 
Hillebrandt, W., et al.\ 2014a, Science, 345, 1162 

\bibitem[Diehl et 
al.(2015)]{2015A&A...574A..72D} Diehl, R., Siegert, T., Hillebrandt, W., et al.\ 2015, \aap, 574, A72 
 

\bibitem[Dong et al.(2014)]{2014arXiv1401.3347D} Dong, S., Katz, B., 
Kushnir, D., \& Prieto, J.~L.\ 2014, arXiv:1401.3347 

\bibitem[Dotani et al.(1987)]{1987Natur.330..230D} Dotani, T., Hayashida, 
K., Inoue, H., Itoh, M., \& Koyama, K.\ 1987, \nat, 330, 230 

\bibitem[Dwarkadas 
\& Chevalier(1998)]{1998ApJ...497..807D} Dwarkadas, V.~V., \& Chevalier, R.~A.\ 1998, \apj, 497, 807 

\bibitem[Fink et al.(2014)]{2014MNRAS.438.1762F} Fink, M., Kromer, M., 
Seitenzahl, I.~R., et al.\ 2014, \mnras, 438, 1762 



\bibitem[Foley et al.(2014)]{2014MNRAS.443.2887F} Foley, R.~J., Fox, O.~D., 
McCully, C., et al.\ 2014, \mnras, 443, 2887 


\bibitem[Fossey et al.(2014)]{2014CBET.3792....1F} Fossey, J., Cooke, B., 
Pollack, G., Wilde, M., 
\& Wright, T.\ 2014, Central Bureau Electronic Telegrams, 3792, 1 

\bibitem[Friesen et al.(2014)]{2014ApJ...792..120F} Friesen, B., Baron, E., 
Wisniewski, J.~P., et al.\ 2014, \apj, 792, 120 



\bibitem[Goobar et al.(2014)]{2014ApJ...784L..12G} Goobar, A., Johansson, 
J., Amanullah, R., et al.\ 2014, \apjl, 784, L12 

%\bibitem[Goobar et al.(2014b)]{2014arXiv1410.1363G} Goobar, A., Kromer, M., 
%Siverd, R., et al.\ 2014b, arXiv:1410.1363 
\bibitem[Goobar et al.(2015)]{2015ApJ...799..106G} Goobar, A., Kromer, M., 
Siverd, R., et al.\ 2015, \apj, 799, 106 


\bibitem[Graham et al.(2015)]{2015MNRAS.446.2073G} Graham, M. L., Foley, R. J.,
 Zheng, W., et al.\ 2015, MNRAS, 446, 2073
 
\bibitem[Greco et al.(2012)]{2012ApJ...757...24G} Greco, J. P., Martini, P., 
\& Thompson, T. A.\ 2012, ApJ, 757, 24
	
\bibitem[Hillebrandt 
\& Niemeyer(2000)]{2000ARA&A..38..191H} Hillebrandt, W., \& Niemeyer, J.~C.\ 2000, \araa, 38, 191 


\bibitem[Hoeflich 
\& Khokhlov(1996)]{1996ApJ...457..500H} Hoeflich, P., \& Khokhlov, A.\ 1996, \apj, 457, 500 

\bibitem[Hoyle 
\& Fowler(1960)]{1960ApJ...132..565H} Hoyle, F., \& Fowler, W.~A.\ 1960, \apj, 132, 565 


\bibitem[Iben 
\& Tutukov(1984)]{1984ApJS...54..335I} Iben, I., Jr., \& Tutukov, A.~V.\ 1984, \apjs, 54, 335 

\bibitem[Imshennik 
\& Dunina-Barkovskaya(2005)]{2005AstL...31..528I} Imshennik, V.~S., \& Dunina-Barkovskaya, N.~V.\ 2005, Astronomy Letters, 31, 528 

 
\bibitem[Isern et 
al.(2013)]{2013A&A...552A..97I} Isern, J., Jean, P., Bravo, E., et al.\ 2013, \aap, 552, A97 


\bibitem[Isern et al.(2014)]{2014ATel.6099....1I} Isern, J., Knoedlseder, 
J., Jean, P., et al.\ 2014, The Astronomer's Telegram, 6099, 1 

\bibitem[Isern et 
al.(2015)]{isern} Isern, J.,  et al.\ 2015, (A\&A, submitted)

\bibitem[Jourdain 
\& Roques(2009)]{2009ApJ...704...17J} Jourdain, E., \& Roques, J.~P.\ 2009, \apj, 704, 17 

\bibitem[Karachentsev 
\& Kashibadze(2006)]{2006Ap.....49....3K} Karachentsev, I.~D., \& Kashibadze, O.~G.\ 2006, Astrophysics, 49, 3 

\bibitem[Kawabata et al.(2014)]{2014ApJ...795L...4K} Kawabata, K.~S., 
Akitaya, H., Yamanaka, M., et al.\ 2014, \apjl, 795, LL4 

\bibitem[Kelly et al.(2014)]{2014ApJ...790....3K} Kelly, P.~L., Fox, O.~D., 
Filippenko, A.~V., et al.\ 2014, \apj, 790, 3 

\bibitem[Kushnir et al.(2013)]{2013ApJ...778L..37K} Kushnir, D., Katz, B., 
Dong, S., Livne, E., \& Fern{\'a}ndez, R.\ 2013, \apjl, 778, LL37 

\bibitem[Kuulkers(2014)]{2014ATel.5835....1K} Kuulkers, E.\ 2014, The 
Astronomer's Telegram, 5835, 1 


\bibitem[Lebrun et al.(2003)]{2003A&A...411L.141L} Lebrun, F., Leray, J.~P., 
 Lavocat, P., et al.\ 2003, \aap, 411, L141 

\bibitem[Leloudas et al.(2009)]{2009A&A...505..265L} Leloudas, G., 
 Stritzinger, M. D., Sollerman, J. et al.\ 2009, A\&A, 505, 265
	
\bibitem[Maeda et al.(2010)]{2010ApJ...708.1703M} Maeda, K., Taubenberger, S., 
Sollerman, J., et al.\ 2010, ApJ, 708, 1703

\bibitem[Maeda et al.(2010)]{2010Natur.466...82M}
Maeda, K., Benetti, S., Stritzinger, M. et al.\ 2010, \nat, 466, 82


\bibitem[Malone et al.(2014)]{2014ApJ...782...11M} Malone, C. M., 
  Nonaka, A., Woosley, S. E., Almgren, A. S., Bell, J. B., Dong, S., 
  \& Zingale, M.\ 2014, \apj, 782, 11 

\bibitem[Margutti et al.(2014)]{2014ApJ...790...52M} Margutti, R., Parrent, 
J., Kamble, A., et al.\ 2014, \apj, 790, 52 


%\bibitem[Marion et al.(2014)]{2014arXiv1405.3970M} Marion, G.~H., Sand, 
%  D.~J., Hsiao, E.~Y., et al.\ 2014, arXiv:1405.3970
\bibitem[Marion et al.(2015)]{2015ApJ...798...39M} Marion, G.~H., Sand, 
D.~J., Hsiao, E.~Y., et al.\ 2015, \apj, 798, 39 
  

\bibitem[Matz et al.(1988)]{1988Natur.331..416M} Matz, S.~M., Share, G.~H., 
Leising, M.~D., Chupp, E.~L., \& Vestrand, W.~T.\ 1988, \nat, 331, 416 

\bibitem[Mazzali et al.(2005)]{2005ApJ...623L..37M}
  Mazzali, P. A., Benetti, S., Altavilla, G. et al.\ 2005, ApJ, 623, L37

\bibitem[Milne et al.(2004)]{2004ApJ...613.1101M} Milne, P.~A., Hungerford, 
A.~L., Fryer, C.~L., et al.\ 2004, \apj, 613, 1101 

\bibitem[Moll et al.(2014)]{2014ApJ...785..105M} Moll, R., Raskin, C., 
Kasen, D., \& Woosley, S.~E.\ 2014, \apj, 785, 105 

\bibitem[Motohara et al.(2006)]{2006ApJ...652L.101M} Motohara, K., 
  Maeda, K., Gerardy, C. et al.\ 2006, \apj, 652, L101

\bibitem[Nadyozhin(1994)]{1994ApJS...92..527N} Nadyozhin, D.~K.\ 1994, 
\apjs, 92, 527 

\bibitem[Nielsen et al.(2014)]{2014MNRAS.442.3400N} Nielsen, M.~T.~B., 
Gilfanov, M., Bogd{\'a}n, {\'A}., Woods, T.~E., 
\& Nelemans, G.\ 2014, \mnras, 442, 3400 


\bibitem[Nomoto et al.(1984)]{1984ApJ...286..644N} Nomoto, K., Thielemann, 
F.-K., \& Yokoi, K.\ 1984, \apj, 286, 644 

\bibitem[Nomoto \& Sugimoto(1977)]{1977PASJ...29..765N} Nomoto, K., 
 \& Sugimoto, D.\ 1977, \pasj, 29, 765 


\bibitem[Nugent et al.(2011)]{2011Natur.480..344N}
  Nugent, P. E., Sullivan, M., Cenko, S. B. et al. 2011, Nature, 480, 344
  
\bibitem[Pakmor et al.(2012)]{2012ApJ...747L..10P} 
   Pakmor, R., Kromer, M., Taubenberger, S., Sim, S. A.,              
	Roepke, F. K., \& Hillebrandt, W.\ 2012, \apj, 747, L10 

%\bibitem[Patat et al.(2014)]{2014arXiv1407.0136P} Patat, F., Taubenberger, 
%  S., Cox, N.~L.~J., et al.\ 2014, arXiv:1407.0136
\bibitem[Patat et 
al.(2015)]{2015A&A...577A..53P} Patat, F., Taubenberger, S., Cox, N.~L.~J., et al.\ 2015, \aap, 577, A53 


  

\bibitem[P{\'e}rez-Torres et al.(2014)]{2014ApJ...792...38P} 
P{\'e}rez-Torres, M.~A., Lundqvist, P., Beswick, R.~J., et al.\ 2014, \apj, 
792, 38 

\bibitem[Piro \& Nakar(2013)]{2013ApJ...769...67P}
   	Piro, A. L., Nakar, E. 2013, ApJ, 769, 67

\bibitem[Roques et 
al.(2003)]{2003A&A...411L..91R} Roques, J.~P., Schanne, S., von Kienlin, A., et al.\ 2003, \aap, 411, L91 

\bibitem[Sazonov et al.(2014)]{2014AstL...40...65S} Sazonov, S.~Y., 
Lutovinov, A.~A., \& Krivonos, R.~A.\ 2014, Astronomy Letters, 40, 65 

\bibitem[Shigeyama et al.(1992)]{1992ApJ...386L..13S} Shigeyama, T., 
  Nomoto, K., Yamaoka, H., \& Thielemann, F.-K.\ 1992, \apjl, 386, L13

\bibitem[Seitenzahl et al.(2013)]{2013MNRAS.429.1156S} Seitenzahl, I.~R., 
Ciaraldi-Schoolmann, F., R{\"o}pke, F.~K., et al.\ 2013, \mnras, 429, 1156 


  

\bibitem[\protect\citeauthoryear{Sturner et 
al.}{2003}]{2003A&A...411L..81S} Sturner S.~J., et al., 2003, A\&A, 411, 
L81 

\bibitem[Sofue(1998)]{1998PASJ...50..227S} Sofue, Y.\ 1998, PASJ, 50, 227

\bibitem[Sunyaev et al.(1987)]{1987Natur.330..227S} Sunyaev, R., Kaniovsky, 
A., Efremov, V., et al.\ 1987, \nat, 330, 227 

\bibitem[Sunyaev et al.(1990)]{1990SvAL...16..171S} Sunyaev, R.~A., 
Kaniovskii, A.~S., Efremov, V.~V., et al.\ 1990, Soviet Astronomy Letters, 
16, 171 


\bibitem[Teegarden et al.(1989)]{1989Natur.339..122T} Teegarden, B.~J., 
Barthelmy, S.~D., Gehrels, N., Tueller, J., 
\& Leventhal, M.\ 1989, \nat, 339, 122 

%\bibitem[Telesco et al.(2014)]{2014arXiv1409.2125T} Telesco, C.~M., 
%H{\"o}flich, P., Li, D., et al.\ 2014, arXiv:1409.2125 
\bibitem[Telesco et al.(2015)]{2015ApJ...798...93T} Telesco, C.~M., 
H{\"o}flich, P., Li, D., et al.\ 2015, \apj, 798, 93 


\bibitem[The 
\& Burrows(2014)]{2014ApJ...786..141T} The, L.-S., \& Burrows, A.\ 2014, \apj, 786, 141 


\bibitem[Ubertini et 
al.(2003)]{2003A&A...411L.131U} Ubertini, P., Lebrun, F., Di Cocco, G., et al.\ 2003, \aap, 411, L131 


\bibitem[Vedrenne et 
al.(2003)]{2003A&A...411L..63V} Vedrenne, G., Roques, J.-P., Sch{\"o}nfelder, V., et al.\ 2003, \aap, 411, L63 

\bibitem[Webbink(1984)]{1984ApJ...277..355W} Webbink, R.~F.\ 1984, \apj, 
277, 355 


\bibitem[Welty et al.(2014)]{2014ApJ...792..106W} Welty, D.~E., Ritchey, 
A.~M., Dahlstrom, J.~A., \& York, D.~G.\ 2014, \apj, 792, 106 

\bibitem[Whelan 
\& Iben(1973)]{1973ApJ...186.1007W} Whelan, J., \& Iben, I., Jr.\ 1973, \apj, 186, 1007 


\bibitem[Winkler et 
al.(2003)]{2003A&A...411L...1W} Winkler, C., Courvoisier, T.~J.-L., Di Cocco, G., et al.\ 2003, \aap, 411, L1 


\bibitem[Woosley et al.(2007)]{2007ApJ...662..487W} Woosley, S.~E., Kasen, 
D., Blinnikov, S., \& Sorokina, E.\ 2007, \apj, 662, 487 


\bibitem[Woosley 
\& Weaver(1991)]{ww91} Woosley, S.~E., \& Weaver, T.~A.\ 1991, in Supernovae (eds Audouze, J., Bludman, S., Mochkovitch, R. \& Zinn-Justin, J.) 63-154 (Elsevier, 1991)  


\bibitem[Woosley 
\& Weaver(1986)]{1986ARA&A..24..205W} Woosley, S.~E., \& Weaver, T.~A.\ 1986, \araa, 24, 205 

\bibitem[Woosley et al.(1981)]{1981CNPPh...9..185W} Woosley, S.~E., 
Axelrod, T.~S., 
\& Weaver, T.~A.\ 1981, Comments Nucl.~Part.~Phys., Vol.~9, p.~185 - 197, 9, 185 

\bibitem[Zheng et al.(2014)]{2014ApJ...783L..24Z} Zheng, W., Shivvers, I., 
Filippenko, A.~V., et al.\ 2014, \apjl, 783, L24 


\end{thebibliography}

\end{document}